\documentclass[aps,prd,reprint,longbibliography,titlepage,nofootinbib]{revtex4-1}

\usepackage{amsthm}
\usepackage{amssymb}   
\usepackage{dsfont}
\usepackage{mathtools} 
\usepackage{hyperref}
\hypersetup{colorlinks=true,linkcolor=blue,urlcolor=blue,citecolor=blue}
\usepackage{accents}
\usepackage{tensor}
\usepackage{xcolor}
\usepackage{graphicx}
\usepackage[cal=boondox]{mathalfa}
\usepackage{lipsum}
\usepackage{relsize}
\usepackage{color}
\usepackage{comment}
\usepackage{nameref}
\usepackage{lineno} 
\newcommand{\etat}{\tilde{\eta}}
\newcommand{\etatu}{\underaccent{\tilde}{\eta}\vspace{0.1cm}\hspace{0.05cm}}

\newcommand{\uac}[1]{\underaccent{\tilde}{#1}}
\newcommand{\uacc}[1]{\uac{\uac{#1}}}
\newcommand{\Q}{\mathcal{Q}}
\newcommand{\h}{h^{\frac{1}{2(n-2)}}}
\newcommand{\ih}{h^{-\frac{1}{2(n-2)}}}
\newcommand{\hs}{h^{\frac{1}{n-2}}}

\begin{document}
	
	\title{Fermions coupled to the Palatini action in $n$ dimensions}

	\author{Jorge Romero\href{https://orcid.org/0000-0001-8258-6647} {\includegraphics[scale=0.05]{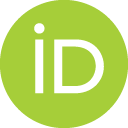}}}
	\email{ljromero@fis.cinvestav.mx}
	\author{Merced Montesinos\href{https://orcid.org/0000-0002-4936-9170} {\includegraphics[scale=0.05]{ORCIDiD_icon128x128.png}}}
	\email[Corresponding author \\]{merced@fis.cinvestav.mx}
	\affiliation{Departamento de F\'{i}sica, Centro de Investigaci\'{o}n y de Estudios Avanzados del Instituto Polit\'{e}cnico Nacional, Avenida Instituto Polit\'{e}cnico Nacional 2508,\\
		San Pedro Zacatenco, 07360 Gustavo Adolfo Madero, Ciudad de M\'{e}xico, Mexico}
	\author{Ricardo Escobedo\href{https://orcid.org/0000-0001-5815-4748} {\includegraphics[scale=0.05]{ORCIDiD_icon128x128.png}}}
	\email[]{ricardo.escobedo@academicos.udg.mx}
	\affiliation{Departamento de F\'{i}sica, Centro Universitario de Ciencias Exactas e Ingenier\'{i}as, Universidad de Guadalajara, Avenida Revoluci\'{o}n 1500, Colonia
		Ol\'{i}mpica, 44430, Guadalajara, Jalisco, Mexico}

	\date{\today}
	
	\begin{abstract}
		We study minimal and nonminimal couplings of fermions to the Palatini action in $n$ dimensions ($n\geq 3$) from the Lagrangian and Hamiltonian viewpoints. The Lagrangian action considered is not, in general, equivalent to the Einstein-Dirac action principle. However, by choosing properly the coupling parameters, it is possible to give a first-order action fully equivalent to the Einstein-Dirac theory in a spacetime of dimension four. By using a suitable parametrization of the vielbein and the connection, the Hamiltonian analysis of the general Lagrangian is given, which involves manifestly Lorentz-covariant phase-space variables, a real noncanonical symplectic structure, and only first-class constraints. Additional Hamiltonian formulations are obtained via symplectomorphisms, one of them involving half-densitized fermions. To confront our results with previous approaches, the time gauge is imposed.  
	\end{abstract}
	
	\maketitle

	
	\section{Introduction}\label{intro}
	General relativity in $n$ dimensions, in the first-order formalism, is given by the Palatini action principle, which depends functionally on the vielbein $e^I$ and the Lorentz connection $\omega^I{}_J$, which are the fundamental independent variables of the theory. This framework is the natural arena to make the coupling of fermions to gravity, which is not possible in the metric formalism of general relativity. When there are no matter fields coupled to gravity, the equation of motion for the connection $\omega^I{}_J$ can be solved to yield $\omega^I{}_J$ as a function of the vielbein and its derivatives, and substituting it into the Palatini action leads to an equivalent second-order action principle for general relativity, which depends only on the vielbein $e^I$. On the other hand, when a fermion field is minimally coupled to the Palatini action, the theory is not equivalent to the Einstein-Dirac theory because of the coupling of the Lorentz connection to the fermion field (see, for instance, Refs.~\cite{Kibble2,Hehl7607} for a spacetime of dimension four). 
	
	In the context of an $n$-dimensional spacetime, the Hamiltonian analysis of fermions minimally coupled to gravity in the first-order formalism has been studied in Ref.~\cite{Bodendorfer1301d}. The Hamiltonian formulation derived there relies on the time gauge. Such a gauge fixing simplifies the handling of the second-class constraints that emerge during the usual Hamiltonian analysis, but it breaks the local Lorentz symmetry in the process. Since the local Lorentz symmetry is one of the fundamental symmetries of nature that is also required to make the coupling of fermions to gravity at the Lagrangian level, it is essential to maintain it during the Hamiltonian analysis to get a deeper understanding
	of the gravity-fermion interaction.
	
	Therefore, in this work, we study the coupling of fermions to the $n$-dimensional Palatini action ($n \geq 3$) in the Hamiltonian formalism without spoiling the local Lorentz invariance. Moreover, to avoid the introduction of second-class constraints in the Hamiltonian analysis---and the complications they imply~\cite{Bodendorfer1301a}---we follow the method presented in Ref.~\cite{Montesinos2001} where authors get the Hamiltonian formulation of the $n$-dimensional Palatini action from scratch by making a suitable parametrization of the vielbein $e^I$ and the connection $\omega^I{}_J$ (see also Ref.~\cite{Montesinos2004a} where the Hamiltonian analysis of the Holst action is performed following the same procedure). An advantage of the approach of Refs.~\cite{Montesinos2001, Montesinos2004a} is that it naturally allows us to identify the manifestly Lorentz-covariant phase-space variables of the theory and, after eliminating the auxiliary fields from the action using their own equations of motion, the Hamiltonian formulation formed solely by first-class constraints easily follows, which simplifies considerably the analysis. This approach has  also been used to study the coupling of fermions to the Holst action~\cite{Romero2106}.
	
	We begin our analysis in Sec.~\ref{Sec_Lagran}, where we present the first-order action principle for a fermion field coupled to the Palatini action in $n$ dimensions used throughout the manuscript. The coupling of the fermion field is generically nonminimal, but it also includes the minimal coupling as a particular case. We eliminate $\omega^I{}_J$ from the action principle using its equation of motion and obtain the equivalent second-order action principle, which turns out to be different from the Einstein-Dirac theory in the generic case. However, we show that a particular choice of the coupling parameters in the first-order Lagrangian action in four dimensions {\it is equivalent} to the Einstein-Dirac action principle plus a boundary term. Next, in Sec.~\ref{Sec_HA}, the Hamiltonian analysis of the general Lagrangian is performed straightforwardly. In Sec.~\ref{Sec_2more}, we present two additional Hamiltonian formulations; one of which is obtained through a symplectomorphism while the other employs half-densitized fermions, which simplifies even more the constraints. For the sake of completeness, we impose the gauge fixing known as time gauge in Sec.~\ref{Sec_tg}, and compare some of our results with those obtained in Ref.~\cite{Bodendorfer1301d}. We finish the paper by making some remarks in Sec.~\ref{Sec_concl}. Our notation and conventions are collected in the Appendices~\ref{ApendiceA}--\ref{Ap_maps}. Further details of the Hamiltonian formulations when the spacetime has dimensions three and four are given in the Appendices~\ref{Ap_3d} and~\ref{Appendixn=4}, respectively. 
	
	\section{Lagrangian analysis}\label{Sec_Lagran}
	
	\subsection{The action principle}
	The gravitational field is given by the $n$-dimensional Palatini---also known as Einstein-Cartan---action
	\begin{equation}
		\label{S_P}
		S_{P}[e, \omega] = \kappa \int_{M} \left[ \star \left( e^{I} \wedge e^{J} \right) \wedge F_{IJ} - 2 \Lambda \rho \right],
	\end{equation}	
	where $\kappa = (16\pi G)^{-1}$ modulates the strength of gravity, $G$ is Newton's gravitational constant, $F^{I}{}_{J} := d \omega^{I}{}_{J} + \omega^{I}{}_{K} \wedge \omega^{K}{}_{J}$ is the curvature of the $SO(n-1,1)$ connection $\omega^{I}{}_{J}$, $\rho:= (1/n!) \epsilon_{I_{1} \ldots I_{n}} e^{I_{1}} \wedge \cdots \wedge e^{I_{n}} $ is the volume form, $\Lambda$ is the cosmological constant, and $\star$ stands for the Hodge dual (see Appendix~\ref{ApendiceA} for more details).
	
	The fermion field $\psi$, coupled to gravity, is given by the action 
	\begin{eqnarray}
		\label{S_F}
		S_{F}[e, \omega, \psi, \bar{\psi}] &:=& \int_{M} \bigg[  \frac{1}{2} \Big( \bar{\psi} \gamma^{I} E D \psi - \overline{D \psi} \gamma^{I} E^{\dagger} \psi \Big) \wedge \star e_{I} \notag \\
		& & - m \bar{\psi} \psi \rho \bigg],
	\end{eqnarray}
	where $\bar{\psi} = \mathrm{i} \psi^{\dagger} \gamma^{0}$, $\gamma^{I}$ are the Dirac matrices, $m$ is the mass of $\psi$, $D$ stands for the covariant derivative with respect to $\omega^{I}{}_{J}$ [see~\eqref{Dp} and \eqref{Dbp}], and $E$ is the coupling matrix defined by
	\begin{equation}
		\label{E}
		E := \left\lbrace  	
		\begin{array}{ll}
			(1 + \mathrm{i} \theta ) \mathds{1} - \mathrm{i} \xi \Gamma, \quad & \mbox{ if $n$ is even} \\ 
			(1 + \mathrm{i} \theta ) \mathds{1},  & \mbox{ if $n$ is odd}
		\end{array}  \right. ,
	\end{equation}
	with $\theta$ and $\xi$ being dimensionless 
	real parameters and $\Gamma$ being the chirality matrix \eqref{CM}. The coupling matrix $E$, $E + E^{\dagger} = 2 \mathds{1}$, involves minimal and nonminimal couplings depending on the values of the parameters. The minimal coupling is when $E=\mathds{1}$, which amounts to set $\theta=\xi=0$. Note that if $n$ is odd, then $\Gamma$ is proportional to $\mathds{1}$, and thus it is already considered in $E$.
	
	It is remarkable that when gravity is turned off, the action principle~\eqref{S_F} leads to the Dirac equation with $m\neq 0$ in an $n$-dimensional Minkowski spacetime for any generic form of the coupling matrix $E$ given by~\eqref{E} (see Appendix~\ref{Ap_Mink}). Thus, the action~\eqref{S_F} has the correct limit when there is no gravity.
	
	In this paper we are interested in the coupling of fermions to general relativity. Therefore, the theory we are going to study is given by the action principle 
	\begin{equation}
		\label{S}
		S[e, \omega, \psi, \bar{\psi}] := S_{P}[e, \omega]  + S_{F}[e, \omega, \psi, \bar{\psi}],
	\end{equation}
	which generalizes the one considered in Ref.~\cite{Bodendorfer1301d}, where authors study only the minimal coupling ($E=\mathds{1}$). 
	
	\subsection{Second-order action} \label{ss_second_order}
	Before performing the Hamiltonian analysis of the first-order action~\eqref{S} and to better understand the nature of the coupling of fermions to gravity, we eliminate the connection $\omega^I{}_J$ from the action principle~\eqref{S} using its equation of motion to get the equivalent second-order action principle, so we can make some remarks regarding the coupling of the fermion field to gravity in both first-order and second-order formalisms.  
	
	The variation of the action~\eqref{S} with respect to the connection $\omega^{I}{}_{J}$ gives the equations of motion
	\begin{eqnarray}
		\label{eom_w} 
		&&\kappa D \left[ \star (e^{I} \wedge e^{J} ) \right] + \dfrac{1}{4} \bigg[  \eta^{K[I}\bar{\psi} \gamma^{J]} (E - E^{\dagger}) \psi  \notag \\
		&&+  \bar{\psi}\{\gamma^{K}, \sigma^{IJ} \} \psi \bigg] \star e_{K} = 0,
	\end{eqnarray}
	where we made use of the fact that $E + E^{\dagger} = 2 \mathds{1}$ and \eqref{anti_comm}. 
	
	The equation of motion~\eqref{eom_w} can be rewritten in the form
	\begin{equation}
		\label{De}
		De^{I} := de^{I} + \omega^{I}{}_{J} \wedge e^{J} = T^{I} ,
	\end{equation}
	where $T^{I}$ is the torsion given by
	\begin{eqnarray}
		T^{I} &:=& \dfrac{1}{8 \kappa} \bigg[ \dfrac{1}{n-2} \bar{\psi} \gamma_{J} (E- E^{\dagger}) \psi e^{I} \wedge e^{J} \notag \\
		&& -  \bar{\psi} \{\gamma^{I}, \sigma_{JK}\} \psi e^{J} \wedge e^{K} \bigg]. 
	\end{eqnarray}
	The solution for $\omega^I{}_J$ is
	\begin{equation}
		\label{sol_w}
		\omega^{I}{}_{J} = \Omega^{I}{}_{J} + C^{I}{}_{J},
	\end{equation}
	where $\Omega^{I}{}_{J} =- \Omega_J{}^I$ is the torsion-free spin connection ($de^{I} + \Omega^{I}{}_{J} \wedge e^{J}  = 0$) and $C^{I}{}_{J} = - C_{J}{}^{I}$ is the contorsion 1-form
	\begin{equation}
		\label{CIJ}
		C_{IJ} := \dfrac{1}{8 \kappa} \left[ \dfrac{2}{n-2} \bar{\psi} \gamma_{[J} (E- E^{\dagger}) \psi e_{I]} +  \bar{\psi} \{\gamma_{K}, \sigma_{IJ}\} \psi  e^{K} \right].
	\end{equation} 
	The contorsion and the torsion are related by $T^{I} = C^{I}{}_{J} \wedge e^{J}$.

	Due to the fact $\omega^{I}{}_{J}$ has been solved using its equation of motion, it is an auxiliary field~\cite{HennBook}. Next, we substitute the solution for the connection \eqref{sol_w} into the action~\eqref{S} and obtain, using~\eqref{g3} and after some algebra, the equivalent second-order action principle
	\begin{eqnarray}\label{2nd_order}
		S_{\mbox{eff}} [e,\psi,\bar{\psi}] &:=& \kappa \int_{M} 
		\left[ \star \left( e^{I} \wedge e^{J} \right) \wedge {\mathcal R}_{IJ} - 2 \Lambda \rho \right] \nonumber\\
		&& +  \int_{M} \bigg[ \dfrac{1}{2} \Big(\bar{\psi} \gamma^{I} D_{\Omega} \psi  - \overline{D_{\Omega} \psi} \gamma^{I} \psi \Big) \wedge \star e_{I}  \nonumber\\
		&& - m \bar{\psi} \psi \rho \bigg] + S_{\mbox{int}}[e,\psi, \bar{\psi}] \notag \\
		&&- \dfrac{1}{4(n-2)}  \int_{\partial M} \bar{\psi} \gamma^{I} ( E - E^{\dagger}) \psi \star e_{I}, \notag \\
	\end{eqnarray}
	where ${\mathcal R}^I{}_J$ is the curvature of $\Omega^I{}_J$, ${\mathcal R}^I{}_J = d \Omega^I{}_J + \Omega^I{}_K \wedge \Omega^K{}_J$, and the covariant derivatives of $\psi$ and $\bar{\psi}$ are given by 
	\begin{subequations}
		\begin{eqnarray}
			D_{\Omega} \psi &:= & d \psi + \frac{1}{2} \Omega_{IJ} \sigma^{IJ} \psi, \\ 
			\overline{D_{\Omega} \psi} &:= &  d \bar{\psi} - \frac{1}{2} \Omega_{IJ} \bar{\psi} \sigma^{IJ}.
		\end{eqnarray}
	\end{subequations}

	A relevant aspect of the second-order Lagrangian formulation~\eqref{2nd_order} is the presence of the interaction term
	\begin{widetext}
		\begin{eqnarray}
			\label{S_int}
			S_{\mbox{int}}[e,\psi, \bar{\psi}] &:=& \dfrac{1}{64 \kappa} \int_{M} \bigg\lbrace \left(\dfrac{n-1}{n-2}\right) \left[ \bar{\psi} \gamma^{I} (E - E^{\dagger}) \psi \right] \left[ \bar{\psi} \gamma_{I} (E - E^{\dagger}) \psi \right]  + \left[ \bar{\psi} \{ \gamma^{I}, \sigma^{JK} \} \psi \right] \left[ \bar{\psi} \{ \gamma_{I}, \sigma_{JK} \} \psi \right] \bigg\rbrace \rho.
		\end{eqnarray}
	\end{widetext}
	Therefore, due to the interaction term $S_{\mbox{int}}$, the resulting second-order action~\eqref{2nd_order} is generically different from the Einstein-Dirac theory, unless the interaction term vanishes. Note that the last term in~\eqref{S_int} corresponds to the well-known interaction term predicted by the Einstein-Cartan theory (see, for instance, Ref.~\cite{Hehl7607}).
	
	However, in a four-dimensional spacetime it is possible to choose the coupling parameters in the first-order action~\eqref{S} in such a way that the resulting second-order action~\eqref{2nd_order} is precisely the Einstein-Dirac theory. This is shown next.
	
	\subsubsection{Four-dimensional spacetime}
	\label{ss_second_order_4d}
	If $n=4$, then we have the result for the anticommutator (see Appendix~\ref{ApendiceA})
	\begin{eqnarray}
		\label{id_4d}
		\{ \gamma^{I}, \sigma^{JK} \} = \mathrm{i} \epsilon^{IJKL} \Gamma \gamma_{L}.
	\end{eqnarray}
	Using this, the fact that $E - E^{\dagger}= 2 \mathrm{i} \left( \theta \mathds{1} - \xi \Gamma \right)$, and taking into account the definition of the real vector $V^I$ and axial $A^I$ currents given by 
	\begin{subequations}
		\begin{eqnarray}
			\label{Vc_def}
			V^{I} &:=& \mathrm{i} \bar{\psi} \gamma^{I} \psi, \\
			\label{Ac_def}
			A^{I} &:=& \mathrm{i} \bar{\psi} \Gamma \gamma^{I} \psi,
		\end{eqnarray}
	\end{subequations}
	the interaction term~\eqref{S_int} acquires the form
	\begin{equation}
		\label{S_int_4d_AV}
		S_{\mbox{int}} = \dfrac{3}{32 \kappa} \int_{M} \bigg[ \theta^{2} V_{I} V^{I} + 2\theta \xi V_{I} A^{I} + \left( \xi^{2} - 1 \right) A_{I} A^{I} \bigg] \rho.
	\end{equation}	
	It is clear that the interaction term is not invariant under the parity transformation due to the middle term in~\eqref{S_int_4d_AV}. However, for the couplings when $\theta=0$ or $\xi=0$, the middle term vanishes, and the interaction term is invariant under parity transformations.\footnote{The same holds for any even dimension. This conclusion comes from writing~\eqref{S_int} in terms of the axial and vector currents for even dimensions.} Note that any of these two choices is not the Einstein-Dirac theory. 
	
	Furthermore, even if we take both  $\theta=0=\xi$, the resulting theory is also not the Einstein-Dirac theory because of the presence of the axial-axial term in~\eqref{S_int_4d_AV}, i.e., the minimal coupling ($E=\mathds{1}$) in the first-order formalism~\eqref{S} is not equivalent to the Einstein-Dirac theory.  
	
	Nevertheless, if we consider the particular choice $\theta=0$ and $\xi=\pm 1 =:\tau$, the interaction term vanishes
	\begin{equation}
		\label{S_int_4d_0}
		S_{\mbox{int}} = 0.
	\end{equation}	
	Thus, in a four-dimensional spacetime, the first-order action~\eqref{S} with nonminimal coupling matrix $E= \mathds{1} - \tau \mathrm{i} \Gamma$ is ---eliminating the connection $\omega^I{}_J$ from~\eqref{S} using its equation of motion--- equivalent to the Einstein-Dirac action plus a boundary term
	\begin{eqnarray}
		S_{\mbox{eff}} &=& \kappa \int_{M} \left[ \star \left( e^{I} \wedge e^{J} \right) \wedge {\mathcal R}_{IJ} - 2 \Lambda \rho \right] \nonumber\\
		&& +  \int_{M} \bigg[ \dfrac{1}{2} \Big(\bar{\psi} \gamma^{I} D_{\Omega} \psi - \overline{D_{\Omega} \psi} \gamma^{I} \psi \Big) \wedge \star e_{I} - m \bar{\psi} \psi \rho \bigg] \nonumber\\
		&& - \dfrac{\tau}{4}  \int_{\partial M} A^{I} \star e_{I}. \notag \\
		&& 
	\end{eqnarray}
	Therefore, the usual belief that the first-order formalism of fermions coupled to gravity is intrinsically different from the second-order formalism given by the Einstein-Dirac theory is not true. As we have shown, it is possible to make them equivalent to each other by choosing a particular nonminimal coupling in the first-order formalism\footnote{An analogous situation happens for the nonminimal coupling of fermions to the Holst action. By making the particular choice of the parameters in the coupling matrix, $\theta=0$ and $\xi=(1/\gamma) (-1 \pm \sqrt{1+\gamma^{2}})$, where $\gamma$ is the Barbero-Immirzi parameter, the interaction term, given in Eq.~(21) of Ref.~\cite{Romero2106}, vanishes $S_{\mbox{int}}=0$.}.

	\section{Hamiltonian analysis}
	\label{Sec_HA}
	Dirac's approach to Hamiltonian systems calls for the definition of the momenta canonically conjugate to all configuration variables~\cite{dirac1964lectures}, enlarging in this way the phase space of the theory under consideration, which is cumbersome most of the times. The method requires us to also evolve the primary constraints and find all the constraints, which must be classified into first class and second class. On the other hand, in first-order gravity for $n >4$, the issue of the second-class constraints becomes still more complicated because they are reducible~\cite{Bodendorfer1301a}, which must be handled somehow~\cite{Bodendorfer1301b}. If, additionally, the coupling of fermions to general relativity is considered, it is expected that the analysis becomes worse. 
	
	Thus, to avoid these issues, we follow the method developed in Refs.~\cite{Montesinos2001, Montesinos2004a}, which consists in a three-step algorithm, to neatly arrive at the Hamiltonian formulations of the $n$-dimensional Palatini and Holst actions involving only first-class constraints and manifestly Lorentz-covariant phase-space variables. This method has also been successfully applied to get the Hamiltonian formulation of fermions coupled to the Holst action~\cite{Romero2106}. 
	
	In the first step of the approach, we parametrize the orthonormal frame of 1-forms (vielbein) $e^I$, adapting it to the geometry of the spacetime foliation. In the second step, we use the parametrization of the connection $\omega^{I}{}_{J}$ naturally induced by the parametrization of the vielbein, which leads to the phase-space variables of the theory. Finally, in the third step, we get rid off the auxiliary fields that do not play a dynamical role in the Hamiltonian formulation by eliminating them from the action principle by using their own equations of motion. All of this is done in what follows. 
	
	\subsection{Parametrization of the vielbein}
	We assume that the spacetime manifold $M$ is diffeomorphic to $\mathbb{R} \times \Sigma$, with $\Sigma$ being a ($n-1$)-dimensional spacelike hypersurface without boundary. Then, we foliate the spacetime with hypersurfaces $\Sigma_{t}$ for every $t\in \mathbb{R}$, and each $\Sigma_{t}$ is diffeomorphic to $\Sigma$. Thus, adapted to the foliation, the local coordinates $(x^{\mu}) = (t, x^{a})$ label the points on $\mathbb{R}$ and $\Sigma$, respectively.
	
	Thus, adapted to the foliation, we write the orthonormal frame of 1-forms and the connection as
	\begin{subequations}
		\begin{eqnarray}
			\label{ed}
			e^{I} &=& e_{0}{}^{I} dt + e_{a}{}^{I} dx^{a}, \\
			\label{wd}
			\omega^{I}{}_{J} &=& \omega_{0}{}^{I}{}_{J} dt + \omega_{a}{}^{I}{}_{J} dx^{a}.
		\end{eqnarray}
	\end{subequations}
	We parametrize the $n^{2}$ components $e_{\mu}{}^{I}$ in terms of the tensor density $\tilde{\Pi}^{aI}$ plus the usual lapse function $N$ and the shift vector $N^{a}$ as
	\begin{subequations}
		\begin{eqnarray}
			\label{ep1}
			e_{0}{}^{I} & = & N n^{I} + N^{a} \h \uac{\uac{h}}_{ab} \tilde{\Pi}^{bI}, \\
			\label{ep2}
			e_{a}{}^{I} & = & \h \uac{\uac{h}}_{ab} \tilde{\Pi}^{bI},
		\end{eqnarray} 
	\end{subequations}
	where
	\begin{equation} 
		\label{n}
		n_{I} := \frac{1}{(n-1)! \sqrt{h}} \epsilon_{IJ_{1} \ldots J_{n-1}} \uac{\eta}_{a_{1} \ldots a_{n-1}} \tilde{\Pi}^{a_{1} J_{1}} \cdots \tilde{\Pi}^{a_{n-1} J_{n-1}} 
	\end{equation}
	is an internal vector orthogonal to $\Sigma$ that satisfies $n_{I} n^{I} = -1 $ and $n_{I} \tilde{\Pi}^{aI} = 0$; $\uac{\uac{h}}_{ab}$ is the densitized metric on $\Sigma$ whose 
	inverse is given by $\tilde{\tilde{h}}^{ab} := \tilde{\Pi}^{aI} \tilde{\Pi}^{b}{}_{I}$, and $h := \det (\tilde{\tilde{h}}^{ab})$ is a tensor density of weight $2(n-2)$. The maps \eqref{ep1} and \eqref{ep2} are invertible, see Appendix~\ref{Ap_maps} for the supplementary maps.
	
	Continuing with the analysis, we use the decomposition of $e^{I}$ and $\omega^{I}{}_{J}$ given in \eqref{ed} and \eqref{wd} together with the parametrization~\eqref{ep1} and~\eqref{ep2}, and we substitute these expressions into the action~\eqref{S} and obtain
	\begin{eqnarray}
		\label{S_std}
		S & = & \int_{\mathbb{R} \times \Sigma} \!\!\!\!\!\!\!\! dt d^{n-1}x \Big[  - 2 \kappa \tilde{\Pi}^{aI} n^{J} \dot{\omega}_{aIJ} + \dfrac{1}{2} \h n_{I} \bar{\psi} \gamma^{I} E \dot{\psi} \notag \\
		& &  - \dfrac{1}{2} \h n_{I} \dot{\bar{\psi}} \gamma^{I} E^{\dagger} \psi + \omega_{0IJ} \tilde{\mathcal{G}}^{IJ} - N^{a} \tilde{\mathcal{V}}_{a} - \uac{N} \tilde{\tilde{\mathcal{S}}} \Big], \notag \\
	\end{eqnarray}
	where $dtd^{n-1}x := dt \wedge dx^{1} \wedge \cdots \wedge dx^{n-1}$, the dot over the corresponding field denotes $\partial_t$, $\uac{N} := \ih N$, and
	\begin{widetext}
		\begin{subequations}\label{std}
			\begin{eqnarray}
				\tilde{\mathcal{G}}^{IJ} & := &  2\kappa \Big[ -  \partial_{a} \left( \tilde{\Pi}^{a[I} n^{J]} \right) +   \omega_{a}{}^{[I}{}_{K} \tilde{\Pi}^{|a|J]}n^{K} -  \omega_{a}{}^{[I}{}_{K} n^{J]} \tilde{\Pi}^{aK} \Big]  + \dfrac{1}{4} \h n_{K} \bar{\psi} \left( \gamma^{K} \sigma^{IJ} E + \sigma^{IJ} \gamma^{K}  E^{\dagger} \right) \psi, \\
				\tilde{\mathcal{V}}_{a} &:=& -2 \kappa \tilde{\Pi}^{bI} n^{J} F_{abIJ} + \frac{1}{2} \h n_{I} \Big( \bar{\psi} \gamma^{I} E D_{a} \psi - \overline{D_{a} \psi} \gamma^{I} E^{\dagger} \psi \Big) \\
				& = & 2 \kappa \Big[ -\tilde{\Pi}^{bI} n^{J} \partial_{a} \omega_{bIJ} + \partial_{b} \Big( \omega_{aIJ} \tilde{\Pi}^{bI} n^{J} \Big) \Big] + \frac{1}{2} \h n_{I}  \left( \bar{\psi} \gamma^{I} E \partial_{a} \psi - \partial_{a} \bar{\psi} \gamma^{I} E^{\dagger} \psi \right)  + \omega_{aIJ}\tilde{\mathcal{G}}^{IJ},  \\
				\tilde{\tilde{\mathcal{S}}} & := & \kappa \tilde{\Pi}^{aI} \tilde{\Pi}^{bJ} F_{abIJ} + \frac{1}{2} h^{\frac{1}{2(n-2)}} \tilde{\Pi}^{aI} \Big( \bar{\psi} \gamma_{I} E D_{a} \psi - \overline{D_{a} \psi} \gamma_{I} E^{\dagger} \psi \Big) - \hs \left( 2 \kappa \Lambda +  m   \bar{\psi} \psi \right),
			\end{eqnarray}
		\end{subequations}
	\end{widetext}
	with
	\begin{eqnarray}
		\label{D_def}
		D_{a} \psi & := & \partial_{a} \psi + \dfrac{1}{2} \omega_{aIJ} \sigma^{IJ} \psi, \\
		\overline{D_{a} \psi} & := & \partial_{a} \bar{\psi} - \dfrac{1}{2} \omega_{aIJ} \bar{\psi} \sigma^{IJ}, \\
		F_{ab}{}^{I}{}_{J} & = & \partial_{a} \omega_{b}{}^{I}{}_{J} - \partial_{b} \omega_{a}{}^{I}{}_{J} + \omega_{a}{}^{I}{}_{K} \omega_{b}{}^{K}{}_{J} \notag \\ 
		&& - \omega_{b}{}^{I}{}_{K} \omega_{a}{}^{K}{}_{J}.
	\end{eqnarray}
	
	Before making the parametrization of the connection, we introduce the covariant derivative $\nabla_{a}$ compatible with $\tilde{\Pi}^{aI}$ 
	\begin{eqnarray}
		\label{CDP}
		\nabla_{a}\tilde{\Pi}^{b I}&:=&  \partial_{a}\tilde{\Pi}^{b I} - \Gamma^{c}{}_{a c}\tilde{\Pi}^{b I} + \Gamma^{b}{}_{a c}\tilde{\Pi}^{c I}  + \Gamma_{a}{}^{I}{}_{J}\tilde{\Pi}^{b J}  =0. \notag \\
	\end{eqnarray}
	This definition is a set of $n(n-1)^{2}$ equations that uniquely determine the $(1/2) n(n-1)^{2} + (1/2) n(n-1)^{2} $ connection components of $\Gamma_{a}{}^{I}{}_{J} = -\Gamma_{aJ}{}^{I}$ and $\Gamma^{a}{}_{bc} = \Gamma^{a}{}_{cb}$. Also, we define the curvature of $\Gamma_{a}{}^{I}{}_{J}$ as $R_{ab}{}^{I}{}_{J} := \partial_{a} \Gamma_{b}{}^{I}{}_{J} - \partial_{b} \Gamma_{a}{}^{I}{}_{J} + \Gamma_{a}{}^{I}{}_{K} \Gamma_{b}{}^{K}{}_{J} - \Gamma_{b}{}^{I}{}_{K} \Gamma_{a}{}^{K}{}_{J}$.

	\subsection{Parametrization of the connection}
	To introduce the suitable parametrization of the connection, we focus our attention on the first term of the action~\eqref{S_std}. We rewrite it as
	\begin{eqnarray}
		\label{sp_Q}
		- 2 \tilde{\Pi}^{aI} n^{J} \dot{\omega}_{aIJ} &=& -2\tilde{\Pi}^{aI} n^{J} \partial_t \left ( \omega_{aIJ} - \Gamma_{aIJ} + \Gamma_{aIJ} \right ) \notag \\
		&=& -2\tilde{\Pi}^{aI} n^{J} \partial_t \left ( \omega_{aIJ} - \Gamma_{aIJ} \right ) \notag \\
		&& -2 \partial_a \left ( n_I \partial_t \tilde{\Pi}^{aI} \right ) \notag \\
		&=&	2 \tilde{\Pi}^{a I}\partial_{t}  \left [ W_{a}{}^{b}{}_{I J K} \left ( \omega_{b}{}^{J K} - \Gamma_{b}{}^{J K} \right ) \right ] \notag \\
		&& -2 \partial_a \left ( n_I \partial_t \tilde{\Pi}^{aI} \right ),
	\end{eqnarray}
	where $W_{a}{}^{b}{}_{IJK} = -W_{a}{}^{b}{}_{IKJ}$ is given by
	\begin{equation}
		\label{W}
		W_{a}{}^{b}{}_{IJK}:= - \delta_{a}^{b} \eta_{I [J}n_{K]} -  \uac{\uac{h}}_{a c} n_{I} \tilde{\Pi}^{c}{}_{[J}\tilde{\Pi}^{b}{}_{K]}.
	\end{equation}
	Therefore, from~\eqref{sp_Q} it is natural to define the $n(n-1)$ phase-space variables 
	\begin{equation}
		\label{Q_def}
		Q_{a I} := W_{a}{}^{b}{}_{I J K} \left( \omega_{b}{}^{J K} - \Gamma_{b}{}^{JK} \right).
	\end{equation}
	Thus, the projector $W_{a}{}^{b}{}_{IJK}$ singles out the dynamic components of $\omega_{a}{}^{I}{}_{J}$. Hence, neglecting the boundary term, the gravitational part of our Hamiltonian formalism is described by the phase-space variables $(Q_{aI}, \tilde{\Pi}^{aI})$. To express the theory in terms of them, we invert \eqref{Q_def}, which is a system of $n(n-1)$ linear equations for $n(n-1)^{2}/2$ unknowns $\omega_{a}{}^{I}{}_{J}$. Therefore, the solution for $\omega_{a}{}^{I}{}_{J}$ must involve $n(n-1)^{2}/2-n(n-1)=n(n-1)(n-3)/2$ free variables. We call these variables $\uac{\uac{u}}_{abc}$, and they satisfy $\uac{\uac{u}}_{abc}= - \uac{\uac{u}}_{acb}$ and the trace condition $\tilde{\tilde{h}}^{ab} \uac{\uac{u}}_{abc} =0$; both conditions account for the correct number of independent variables contained in $\uac{\uac{u}}_{abc}$.
	
	From \eqref{Q_def}, the solution for $\omega_{a}{}^{I}{}_{J}$ is
	\begin{equation}
		\label{wQ}
		\omega_{aIJ} = M_{a}{}^{b}{}_{IJK} Q_{b}{}^{K} + \tilde{\tilde{N}}_a{}^{bcd}{}_{IJ}\uac{\uac{u}}_{bcd} + \Gamma_{aIJ},
	\end{equation}
	where $M_{a}{}^{b}{}_{IJK} = -M_{a}{}^{b}{}_{JIK}$ and $\tilde{\tilde{N}}_a{}^{bcd}{}_{IJ}=-\tilde{\tilde{N}}_a{}^{bcd}{}_{JI}=-\tilde{\tilde{N}}_a{}^{bdc}{}_{IJ}$ are functions of $\tilde{\Pi}^{aI}$ that are given in \eqref{M} and \eqref{N}, respectively.

	Now that we have the parametrization for the spatial components of the connection, we substitute~\eqref{wQ} into the action~\eqref{S_std} and obtain
	\begin{eqnarray}
		\label{S_GQ}
		S & = & \int_{\mathbb{R} \times \Sigma} dtd^{n-1}x \bigg[  2 \kappa \tilde{\Pi}^{aI} \dot{Q}_{aI} + \dfrac{1}{2} \h n_{I} \Big( \bar{\psi} \gamma^{I} E \dot{\psi}  \notag \\
		&&-  \dot{\bar{\psi}} \gamma^{I} E^{\dagger} \psi \Big) + \omega_{0IJ} \tilde{\mathcal{G}}^{IJ} - N^{a} \tilde{\mathcal{V}}_{a} - \uac{N} \tilde{\tilde{\mathcal{S}}} \bigg],
	\end{eqnarray}
	with 
	\begin{widetext}
		\begin{subequations} 
			\label{const_GQ}
			\begin{eqnarray}
				\tilde{\mathcal{G}}^{IJ} & = & 2 \kappa \tilde{\Pi}^{a[I} Q_{a}{}^{J]}
				+ \dfrac{1}{4} \h n_{K} \bar{\psi} \left( \gamma^{K} \sigma^{IJ} E + \sigma^{IJ} \gamma^{K}  E^{\dagger} \right) \psi, \\
				\tilde{\mathcal{V}}_{a} & = & 2 \kappa \Big( 2 \tilde{\Pi}^{bI} \partial_{[a} Q_{b]I} - Q_{aI} \partial_{b} \tilde{\Pi}^{bI} \Big) + \frac{1}{2} \h n_{I}  \left( \bar{\psi} \gamma^{I} E  \partial_{a} \psi - \partial_{a} \bar{\psi} \gamma^{I} E^{\dagger} \psi \right) \notag \\
				&& + \Big( M_{a}{}^{b}{}_{IJK} Q_{b}{}^{K} + \tilde{\tilde{N}}_a{}^{bcd}{}_{IJ}\uac{\uac{u}}_{bcd} + \Gamma_{aIJ} \Big) \tilde{\mathcal{G}}^{IJ},  \\
				\tilde{\tilde{\mathcal{S}}} & = & \kappa \tilde{\Pi}^{aI} \tilde{\Pi}^{bJ} R_{abIJ} + 2 \kappa \tilde{\Pi}^{a[I} \tilde{\Pi}^{|b|J]} Q_{aI} Q_{bJ} + \dfrac{1}{2} \h \tilde{\Pi}^{aI} \left(\bar{\psi} \gamma_{I}  \nabla_{a} \psi - \overline{\nabla_{a} \psi} \gamma_{I} \psi \right) \notag \\
				&& + \dfrac{1}{2} \h  Q_{aI} n_{J} \tilde{\Pi}^{a}{}_{K} \bar{\psi} \Big( \gamma^{K} \sigma^{IJ} E + \sigma^{IJ} \gamma^{K} E^{\dagger} \Big) \psi + \dfrac{(n-1)\hs}{64 (n-2) \kappa}q_{IJ} \left[ \bar{\psi} \gamma^{I} (E-E^{\dagger}) \psi \right] \left[ \bar{\psi} \gamma^{J} (E-E^{\dagger}) \psi \right] \notag \\
				&& + \uacc{u}_{abc} \left( \kappa \tilde{\tilde{h}}^{db} \tilde{\tilde{h}}^{cf} \tilde{\tilde{h}}^{ea} \uacc{u}_{def} + \dfrac{1}{4} \h \tilde{\Pi}^{a}{}_{I} \tilde{\Pi}^{b}{}_{J} \tilde{\Pi}^{c}{}_{K} \bar{\psi} \{\gamma^{I}, \sigma^{JK}\}  \psi \right) - \hs \left( 2 \kappa \Lambda + m \bar{\psi} \psi  \right) \notag \\
				\label{HGQ}
				&&   + \dfrac{1}{\kappa (n-2)}\tilde{\mathcal{G}}_{IJ} \left[ (n-3) n^{I} n_{K} \tilde{\mathcal{G}}^{JK} + \dfrac{1}{4} \h n^{I}  \bar{\psi} \gamma^{J} (E-E^{\dagger}) \psi \right]   + 2 \tilde{\Pi}^{aI} n^{J} \nabla_{a} \tilde{\mathcal{G}}_{IJ},
			\end{eqnarray}
		\end{subequations}
	where $q_{IJ}  :=  \eta_{IJ} + n_{I} n_{J}$ and the covariant derivatives are given by
	\begin{subequations}
		\begin{eqnarray}
			\label{n_psi}
			\nabla_{a} \psi & := & \partial_{a} \psi + \dfrac{1}{2} \Gamma_{aIJ} \sigma^{IJ} \psi, \\
			\label{n_bpsi}
			\overline{\nabla_{a} \psi} & := & \partial_{a} \bar{\psi} - \dfrac{1}{2} \Gamma_{aIJ} \bar{\psi} \sigma^{IJ},  \\
			\nabla_{a} \tilde{\mathcal{G}}^{IJ} & := & \partial_{a} \tilde{\mathcal{G}}^{IJ} - \Gamma^{b}{}_{ba} \tilde{\mathcal{G}}^{IJ} + \Gamma_{a}{}^{I}{}_{K} \tilde{\mathcal{G}}^{KJ} + \Gamma_{a}{}^{J}{}_{K} \tilde{\mathcal{G}}^{IK}.
		\end{eqnarray}
	\end{subequations}
	
	We simplify the expressions by factoring all the terms involving $\tilde{\mathcal{G}}^{IJ}$ in \eqref{S_GQ}. Thus, after integrating by parts the last term of~\eqref{HGQ}, and redefining the Lagrange multiplier $\omega_{0IJ}$ as 
	\begin{eqnarray}
		\label{lQ}
		\omega_{0IJ} & =: & - \lambda_{IJ} + N^{a} \Big( M_{a}{}^{b}{}_{IJK} Q_{b}{}^{K} + \tilde{\tilde{N}}_a{}^{bcd}{}_{IJ}\uac{\uac{u}}_{bcd} + \Gamma_{aIJ} \Big) - 2 \tilde{\Pi}^{a}{}_{[I} n_{J]} \nabla_{a} \uac{N} \notag \\ 
		&& + \dfrac{\uac{N}}{\kappa (n-2)} \Big[ (n-3) n_{[I} n^{K} \tilde{\mathcal{G}}_{J]K}  + \dfrac{1}{4} \h n_{[I}  \bar{\psi} \gamma_{J]} (E-E^{\dagger}) \psi \Big],  
	\end{eqnarray}
	the action~\eqref{S_GQ} becomes 
	\begin{eqnarray}
		\label{S_Qu}
		S & = & \int_{\mathbb{R} \times \Sigma} dtd^{n-1}x \bigg[  2 \kappa \tilde{\Pi}^{aI} \dot{Q}_{aI} + \dfrac{1}{2} \h n_{I} \Big( \bar{\psi} \gamma^{I} E \dot{\psi} -  \dot{\bar{\psi}} \gamma^{I} E^{\dagger} \psi \Big) - \lambda_{IJ} \tilde{\mathcal{G}}^{IJ} - 2 N^{a} \tilde{\mathcal{D}}_{a} - \uac{N} \tilde{\tilde{\mathcal{Z}}} \bigg],
	\end{eqnarray}
	with 
		\begin{subequations} 
			\label{const_Qu}
			\begin{eqnarray}
				\tilde{\mathcal{G}}^{IJ} & = & 2 \kappa \tilde{\Pi}^{a[I} Q_{a}{}^{J]}
				+ \dfrac{1}{4} \h n_{K} \bar{\psi} \left( \gamma^{K} \sigma^{IJ} E + \sigma^{IJ} \gamma^{K}  E^{\dagger} \right) \psi, \\
				\tilde{\mathcal{D}}_{a} & := & \kappa \Big( 2 \tilde{\Pi}^{bI} \partial_{[a} Q_{b]I} - Q_{aI} \partial_{b} \tilde{\Pi}^{bI} \Big) + \frac{1}{4} \h n_{I}  \left( \bar{\psi} \gamma^{I} E  \partial_{a} \psi - \partial_{a} \bar{\psi} \gamma^{I} E^{\dagger} \psi \right), \\
				\tilde{\tilde{\mathcal{Z}}} & := & \kappa \tilde{\Pi}^{aI} \tilde{\Pi}^{bJ} R_{abIJ} + 2 \kappa \tilde{\Pi}^{a[I} \tilde{\Pi}^{|b|J]} Q_{aI} Q_{bJ} + \dfrac{1}{2} \h \tilde{\Pi}^{aI} \left( \bar{\psi} \gamma_{I} \nabla_{a} \psi - \overline{\nabla_{a} \psi} \gamma_{I} \psi \right) \notag \\
				&& + \dfrac{1}{2} \h  Q_{aI} n_{J} \tilde{\Pi}^{a}{}_{K} \bar{\psi} \Big( \gamma^{K} \sigma^{IJ} E + \sigma^{IJ} \gamma^{K} E^{\dagger} \Big) \psi + \dfrac{(n-1)\hs}{64 (n-2) \kappa}q_{IJ} \left[ \bar{\psi} \gamma^{I} (E-E^{\dagger}) \psi \right] \left[ \bar{\psi} \gamma^{J} (E-E^{\dagger}) \psi \right] \notag \\
				&& + \uacc{u}_{abc} \left[ \kappa \tilde{\tilde{h}}^{db} \tilde{\tilde{h}}^{cf} \tilde{\tilde{h}}^{ea} \uacc{u}_{def} + \dfrac{1}{4} \h \tilde{\Pi}^{a}{}_{I} \tilde{\Pi}^{b}{}_{J} \tilde{\Pi}^{c}{}_{K} \bar{\psi} \{\gamma^{I}, \sigma^{JK}\} \psi \right] - \hs \left( 2 \kappa \Lambda + m \bar{\psi} \psi  \right).			
			\end{eqnarray}
		\end{subequations}

	Until this point, we have mapped the $n^{2}$ components of the orthonormal frame of 1-forms $(e_{\mu}{}^{I}) \mapsto (N, N^{a}, \tilde{\Pi}^{aI})$ and the $n^{2}(n-1)/2$ components of the connection $(\omega_{\mu}{}^{I}{}_{J}) \mapsto (Q_{aI}, \uacc{u}_{abc}, \lambda_{IJ})$. The parametrization of the connection is obviously not unique, since we can define other variables (see, for instance, Ref.~\cite{Montesinos2001} where alternative variables are induced when no boundary term is neglected). However, we have chosen the phase-space variables $(Q_{aI}, \tilde{\Pi}^{aI})$ because they have a clear geometrical meaning; both transform as Lorentz vectors under local $SO(n-1,1)$ transformations.  
	
	\newpage
\end{widetext}

	\subsection{Eliminating the auxiliary fields}
	
	Although it appears that we have reached a Hamiltonian description, this is not so because we still need to handle the variables $\uacc{u}_{abc}$. According to Dirac's method, the definition of the momenta canonically conjugate to $\uacc{u}_{abc}$ is required, which would introduce second-class constraints and would enlarge the phase space again. Furthermore, such second-class constraints must be handled somehow and things become complicated. Therefore, we circumvent Dirac's method, following an alternative way that avoids all of this.

	The variables $\uacc{u}_{abc}$ are auxiliary fields~\cite{HennBook}. In fact, from the variation of the action with respect to $\uacc{u}_{abc}$, we get the equation of motion 
	\begin{eqnarray}
		\label{eom_u}
		&& \dfrac{\uac{N}}{4} \h \tilde{\Pi}^{a}{}_{I} \tilde{\Pi}^{[b}{}_{J} \tilde{\Pi}^{c]}{}_{K} \bar{\psi} \{\gamma^{I}, \sigma^{JK}\}  \psi \nonumber\\
		&& + 2 \kappa \uac{N} \tilde{\tilde{h}}^{d[b} \tilde{\tilde{h}}^{c]f} \tilde{\tilde{h}}^{ea} \uacc{u}_{def}  = 0, 
	\end{eqnarray}
	which can be solved for $\uacc{u}_{abc}$:
	\begin{equation}
		\label{sol_u}
		\uacc{u}_{abc} = \dfrac{1}{8\kappa} \h \uacc{h}_{ad} \uacc{h}_{e[b} \uacc{h}_{c]f} \tilde{\Pi}^{d}{}_{I} \tilde{\Pi}^{e}{}_{J} \tilde{\Pi}^{f}{}_{K} \bar{\psi} \{\gamma^{I}, \sigma^{JK}\} \psi.
	\end{equation}
	Substituting \eqref{sol_u} into the action~\eqref{S_Qu} and simplifying, we obtain
	\begin{eqnarray}
		\label{S_Q}
		S & = & \int_{\mathbb{R} \times \Sigma} dtd^{n-1}x \bigg[  2 \kappa \tilde{\Pi}^{aI} \dot{Q}_{aI} + \dfrac{1}{2} \h n_{I} \Big( \bar{\psi} \gamma^{I} E \dot{\psi}  \notag \\
		&&-  \dot{\bar{\psi}} \gamma^{I} E^{\dagger} \psi \Big) - \lambda_{IJ} \tilde{\mathcal{G}}^{IJ} - 2 N^{a} \tilde{\mathcal{D}}_{a} - \uac{N} \tilde{\tilde{\mathcal{H}}} \bigg],
	\end{eqnarray}
	where the Gauss $\tilde{\mathcal{G}}^{IJ}$, diffeomorphism $\tilde{\mathcal{D}}_{a}$, and Hamiltonian $\tilde{\tilde{\mathcal{H}}}$ constraints are given by
	\begin{widetext}
		\begin{subequations} 
			\label{const_Q}
			\begin{eqnarray}
				\label{const_Q_G}
				\tilde{\mathcal{G}}^{IJ} & = & 2 \kappa \tilde{\Pi}^{a[I} Q_{a}{}^{J]}
				+ \dfrac{1}{4} \h \left[ n^{[I} \bar{\psi} \gamma^{J]} ( E - E^{\dagger}) \psi +  n_{K} \bar{\psi} \{\gamma^{K}, \sigma^{IJ}\} \psi \right], \\
				\tilde{\mathcal{D}}_{a} & = & \kappa \Big( 2 \tilde{\Pi}^{bI} \partial_{[a} Q_{b]I} - Q_{aI} \partial_{b} \tilde{\Pi}^{bI} \Big) + \frac{1}{4} \h n_{I}  \left( \bar{\psi} \gamma^{I} E  \partial_{a} \psi - \partial_{a} \bar{\psi} \gamma^{I} E^{\dagger} \psi \right), \\
				\tilde{\tilde{\mathcal{H}}} & := & \kappa \tilde{\Pi}^{aI} \tilde{\Pi}^{bJ} R_{abIJ} + 2 \kappa \tilde{\Pi}^{a[I} \tilde{\Pi}^{|b|J]} Q_{aI} Q_{bJ} + \dfrac{1}{2} \h \tilde{\Pi}^{aI} \left( \bar{\psi} \gamma_{I} \nabla_{a} \psi - \overline{\nabla_{a} \psi} \gamma_{I} \psi \right) \notag \\
				&& + \dfrac{1}{2} \h n_{I} Q_{aJ} \left[ \dfrac{1}{2}\tilde{\Pi}^{aJ}  \bar{\psi} \gamma^{I}(E - E^{\dagger}) \psi -  \tilde{\Pi}^{a}{}_{K} \bar{\psi} \{\gamma^{I}, \sigma^{JK}\} \psi \right] \notag \\
				&& + \dfrac{\hs}{64 \kappa} \bigg\lbrace \left( \dfrac{n-1}{n-2} \right) q_{IJ} \left[ \bar{\psi} \gamma^{I} (E-E^{\dagger}) \psi \right] \left[ \bar{\psi} \gamma^{J} (E-E^{\dagger}) \psi \right]  +  \left( \bar{\psi} \{\gamma^{I}, \sigma^{JK}\} \psi \right) \left( \bar{\psi} \{\gamma_{I}, \sigma_{JK}\} \psi \right) \notag \\
				\label{const_Q_H}
				&& + 3 n_{I} n^{J}  \left( \bar{\psi} \{\gamma^{I}, \sigma^{KL}\} \psi \right) \left( \bar{\psi} \{\gamma_{J}, \sigma_{KL}\} \psi \right)  \bigg\rbrace  - \hs \left( 2 \kappa \Lambda + m \bar{\psi} \psi  \right).			
			\end{eqnarray}
		\end{subequations}
	\end{widetext}

	The constraints $\tilde{\mathcal{G}}^{IJ}$, $\tilde{\mathcal{D}}_{a}$, and $\tilde{\tilde{\mathcal{H}}}$ are first class, and they generate the gauge symmetries of the theory. The Gauss constraint $\tilde{\mathcal{G}}^{IJ}$ generates the local Lorentz transformations, while $\tilde{\mathcal{D}}_{a}$ and $\tilde{\tilde{\mathcal{H}}}$ generate spacetime diffeomorphisms. We highlight that in the formulation~\eqref{S_Q}, which comes out after integrating the auxiliary fields, the remaining field variables are $(\uac{N}, N^a, \tilde{\Pi}^{aI}, Q_{aI}, \lambda_{IJ}, \psi, \bar{\psi})$, from which $(\uac{N}, N^a, \lambda_{IJ})$ play the role of Lagrange multipliers. Furthermore, the phase-space variables $(Q_{aI}, \tilde{\Pi}^{aI})$ transform as vectors under local Lorentz transformations and as a 1-form and a vector density of weight $+1$ under spatial diffeomorphisms, respectively. It is also worth stressing that the Hamiltonian formulation maintains manifestly and completely the Lorentz invariance and that the full noncanonical symplectic structure is real.

	In the case of the minimal coupling ($E=\mathds{1}$), all the terms involving $(E-E^{\dagger})$ vanish, so this case is easily derived from the above formulation. 
	
	Since we have a different matrix coupling $E$ depending on the spacetime dimension $n$ [see \eqref{E}], we bifurcate our analysis next and explicitly show the relevance of the coupling parameters. 
	
	\subsubsection{Even dimensions}
	In the case when the spacetime dimension is even, the coupling matrix is
	\begin{equation}
		\label{E_even}
		E = (1 + \mathrm{i} \theta ) \mathds{1} - \mathrm{i} \xi \Gamma. 
	\end{equation}
	Thus, using the definitions \eqref{Vc_def} and \eqref{Ac_def}, the Hamiltonian formalism is defined by the action~\eqref{S_Q} with the constraints
	\begin{widetext}
		\begin{subequations} 
			\label{const_Q_e}
			\begin{eqnarray}
				\tilde{\mathcal{G}}^{IJ} & = & 2 \kappa \tilde{\Pi}^{a[I} Q_{a}{}^{J]}
				+ \dfrac{1}{2} \h n^{[I} \left( \theta V^{J]}  + \xi A^{J]} \right) + \dfrac{1}{4} \h  n_{K} \bar{\psi} \{\gamma^{K}, \sigma^{IJ}\} \psi , \\
				\tilde{\mathcal{D}}_{a} & = & \kappa \Big( 2 \tilde{\Pi}^{bI} \partial_{[a} Q_{b]I} - Q_{aI} \partial_{b} \tilde{\Pi}^{bI} \Big) + \frac{1}{4} \h n_{I}  \left[\bar{\psi} \gamma^{I} \partial_{a} \psi - \partial_{a} \bar{\psi} \gamma^{I} \psi + \partial_{a}\left( \theta V^{I} + \xi A^{I} \right) \right], \\
				\tilde{\tilde{\mathcal{H}}} & = & \kappa \tilde{\Pi}^{aI} \tilde{\Pi}^{bJ} R_{abIJ} + 2 \kappa \tilde{\Pi}^{a[I} \tilde{\Pi}^{|b|J]} Q_{aI} Q_{bJ} + \dfrac{1}{2} \h \tilde{\Pi}^{aI} \left( \bar{\psi} \gamma_{I} \nabla_{a} \psi - \overline{\nabla_{a} \psi} \gamma_{I} \psi \right) \notag \\
				&& + \dfrac{1}{2} \h  n_{I} Q_{aJ}  \left[ \tilde{\Pi}^{aJ}  \left( \theta V^{I} + \xi A^{I} \right) -  \tilde{\Pi}^{a}{}_{K} \bar{\psi} \{\gamma^{I}, \sigma^{JK}\} \psi \right] \notag \\
				&& + \dfrac{\hs}{64 \kappa} \bigg[ 4 \left( \dfrac{n-1}{n-2} \right) q_{IJ} \left( \theta^{2} V^{I} V^{J} + \xi^{2} A^{I} A^{J} + 2 \theta \xi V^{I} A^{J} \right) +   \left( \bar{\psi} \{\gamma^{I}, \sigma^{JK}\} \psi \right) \left( \bar{\psi} \{\gamma_{I}, \sigma_{JK}\} \psi \right) \notag \\
				&& + 3 n_{I} n^{J}  \left( \bar{\psi} \{\gamma^{I}, \sigma^{KL}\} \psi \right) \left( \bar{\psi} \{\gamma_{J}, \sigma_{KL}\} \psi \right) \bigg]  - \hs \left( 2 \kappa \Lambda + m \bar{\psi} \psi  \right).			
			\end{eqnarray}
		\end{subequations}

	\subsubsection{Odd dimensions}
	When the spacetime dimension is odd, we consider the coupling matrix
	\begin{equation}
		\label{E_odd}
		E = (1 + \mathrm{i} \theta ) \mathds{1}.
	\end{equation}
	Thus, the Hamiltonian formalism is described by the action~\eqref{S_Q} and the constraints are
		\begin{subequations} 
			\label{const_Q_o}
			\begin{eqnarray}
				\label{const_Q_G_o}
				\tilde{\mathcal{G}}^{IJ} & = & 2 \kappa \tilde{\Pi}^{a[I} Q_{a}{}^{J]}
				+ \dfrac{\theta}{2} \h n^{[I} V^{J]} + \dfrac{1}{4} \h  n_{K} \bar{\psi} \{\gamma^{K}, \sigma^{IJ}\} \psi , \\
				\tilde{\mathcal{D}}_{a} & = & \kappa \Big( 2 \tilde{\Pi}^{bI} \partial_{[a} Q_{b]I} - Q_{aI} \partial_{b} \tilde{\Pi}^{bI} \Big) + \frac{1}{4} \h n_{I}  \left(\bar{\psi} \gamma^{I} \partial_{a} \psi - \partial_{a} \bar{\psi} \gamma^{I} \psi + \theta \partial_{a} V^{I} \right), \\
				\tilde{\tilde{\mathcal{H}}} & = & \kappa \tilde{\Pi}^{aI} \tilde{\Pi}^{bJ} R_{abIJ} + 2 \kappa \tilde{\Pi}^{a[I} \tilde{\Pi}^{|b|J]} Q_{aI} Q_{bJ} + \dfrac{1}{2} \h \tilde{\Pi}^{aI} \left( \bar{\psi} \gamma_{I} \nabla_{a} \psi - \overline{\nabla_{a} \psi} \gamma_{I} \psi \right) \notag \\
				&& + \dfrac{1}{2} \h  n_{I} Q_{aJ}  \left( \theta \tilde{\Pi}^{aJ} V^{I} -  \tilde{\Pi}^{a}{}_{K} \bar{\psi} \{\gamma^{I}, \sigma^{JK}\} \psi \right) + \dfrac{\hs}{64 \kappa} \bigg[ 4 \theta^{2} \left( \dfrac{n-1}{n-2} \right)  q_{IJ} V^{I} V^{J} \notag \\
				&&   +  \left( \bar{\psi} \{\gamma^{I}, \sigma^{JK}\} \psi \right) \left( \bar{\psi} \{\gamma_{I}, \sigma_{JK}\} \psi \right)  + 3 n_{I} n^{J}  \left( \bar{\psi} \{\gamma^{I}, \sigma^{KL}\} \psi \right) \left( \bar{\psi} \{\gamma_{J}, \sigma_{KL}\} \psi \right)  \bigg] \notag \\
				\label{const_Q_H_o}
				&&  - \hs \left( 2 \kappa \Lambda + m \bar{\psi} \psi  \right),
			\end{eqnarray}
		\end{subequations}
	where the vector current is defined in~\eqref{Vc_def}.
	\end{widetext}	
	\section{Alternative Hamiltonian formulations}
	\label{Sec_2more}
	We present two additional Hamiltonian formulations of the action~\eqref{S}, which are easily obtained from the Hamiltonian action~\eqref{S_Q}. The first of these formulations is deduced from a symplectomorphism while the second is gotten employing half-densitized fermion fields\footnote{Although we could explore more Hamiltonian formulations as in Ref.~\cite{Romero2106}, we just consider the ones already mentioned.}.

	\subsection{Hamiltonian formulation through a symplectomorphism}
	We make a symplectomorphism that only changes the variable $Q_{aI}$ to
	\begin{equation}
		\label{Q2q}
		\Q_{aI} = Q_{aI} + W_{a}{}^{b}{}_{IJK} \Gamma_{b}{}^{JK},
	\end{equation}
	leaving $\tilde{\Pi}^{aI}$, $\psi$, and $\bar{\psi}$ unchanged. 
	
	Note that in terms of the original connection variables $\omega_{a}{}^{I}{}_{J}$, $\Q_{aI}$ is given by
	\begin{equation}
		\label{q_def}
		\Q_{aI} := W_{a}{}^{b}{}_{IJK} \omega_{b}{}^{JK},
	\end{equation}
	which can be obtained by simply substituting~\eqref{Q_def} into the right-hand side of~\eqref{Q2q} or, alternatively, from writing the first term of~\eqref{S_std} as
	\begin{eqnarray}
		\label{sp_q}
		- 2 \tilde{\Pi}^{aI} n^{J} \dot{\omega}_{aIJ} &=& 2 \tilde{\Pi}^{a I} \partial_{t}  \left( W_{a}{}^{b}{}_{I J K} \omega_{b}{}^{J K} \right) \notag \\
		&=& 2 \tilde{\Pi}^{a I} \dot{\Q}_{aI},
	\end{eqnarray}
	which also shows that no boundary term, as in \eqref{sp_Q}, arises if we had defined these variables from the very beginning in the Hamiltonian analysis.
	
	In terms of the new variables, the symplectic structure in~\eqref{S_Q} becomes
	\begin{eqnarray}\label{SympEst}
		&&2 \kappa \tilde{\Pi}^{aI} \dot{Q}_{aI} + \dfrac{1}{2} \h n_{I} \Big( \bar{\psi} \gamma^{I} E \dot{\psi} - \dot{\bar{\psi}} \gamma^{I} E^{\dagger} \psi \Big) \notag\\
		&& =  2 \kappa \tilde{\Pi}^{aI} \dot{\Q}_{aI} + \dfrac{1}{2} \h n_{I} \Big( \bar{\psi} \gamma^{I} E \dot{\psi} -  \dot{\bar{\psi}} \gamma^{I} E^{\dagger} \psi \Big) \notag \\
		&& \quad +  2 \partial_a \left ( n_I \partial_t \tilde{\Pi}^{aI} \right ),
	\end{eqnarray}
	which shows that the transformation is indeed a symplectomorphism [note that the boundary term in the last line is the one that is neglected in \eqref{sp_Q}].
	
	Therefore, using~\eqref{SympEst} and neglecting the boundary term, we get
	\begin{eqnarray}
		\label{S_q}
		S & = & \int_{\mathbb{R} \times \Sigma} dtd^{n-1}x \bigg[  2 \kappa \tilde{\Pi}^{aI} \dot{\Q}_{aI} + \dfrac{1}{2} \h n_{I} \Big( \bar{\psi} \gamma^{I} E \dot{\psi} \notag \\
		&&  -  \dot{\bar{\psi}} \gamma^{I} E^{\dagger} \psi \Big) - \lambda_{IJ} \tilde{\mathcal{G}}^{IJ} - 2 N^{a} \tilde{\mathcal{D}}_{a} - \uac{N} \tilde{\tilde{\mathcal{H}}} \bigg],
	\end{eqnarray}
	where the Gauss, diffeomorphism, and Hamiltonian constraints now read
	\begin{widetext}
		\begin{subequations} 
			\label{const_q}
			\begin{eqnarray}
				\label{const_q_G}
				\tilde{\mathcal{G}}^{IJ} & = &  2 \kappa \Big( \tilde{\Pi}^{a[I} \Q_{a}{}^{J]} + \tilde{\Pi}^{a[I} \Gamma_{a}{}^{J]}{}_{K} n^{K} - \tilde{\Pi}^{aK} n^{[I} \Gamma_{a}{}^{J]}{}_{K} \Big) + \dfrac{1}{4} \h \Big[ n^{[I} \bar{\psi} \gamma^{J]} ( E - E^{\dagger}) \psi \notag \\
				&& +  n_{K} \bar{\psi} \{\gamma^{K}, \sigma^{IJ}\} \psi \Big], \\
				\tilde{\mathcal{D}}_{a} & = & \kappa \Big( 2 \tilde{\Pi}^{bI} \partial_{[a} \Q_{b]I} - \Q_{aI} \partial_{b} \tilde{\Pi}^{bI} \Big) + \frac{1}{4} \h n_{I}  \left( \bar{\psi} \gamma^{I} E  \partial_{a} \psi - \partial_{a} \bar{\psi} \gamma^{I} E^{\dagger} \psi \right),\\
				\tilde{\tilde{\mathcal{H}}} & = & \kappa \tilde{\Pi}^{aI} \tilde{\Pi}^{bJ} R_{abIJ} + 2 \kappa \tilde{\Pi}^{a[I} \tilde{\Pi}^{|b|J]} \left( \Q_{aI} \Q_{bJ} + 2 \Q_{aI} \Gamma_{bJK} n^{K} + \Gamma_{aIK} \Gamma_{bJL} n^{K} n^{L} \right) \notag \\
				&& + \dfrac{1}{2} \h \tilde{\Pi}^{aI} \left( \bar{\psi} \gamma_{I} \nabla_{a} \psi - \overline{\nabla_{a} \psi} \gamma_{I} \psi \right) + \dfrac{1}{2} \h  n_{I} \left( \Q_{a J} + \Gamma_{aJL} n^{L} \right)  \bigg[ \dfrac{1}{2}\tilde{\Pi}^{aJ}  \bar{\psi} \gamma^{I}(E - E^{\dagger}) \psi \notag \\
				&&  -  \tilde{\Pi}^{a}{}_{K} \bar{\psi} \{\gamma^{I}, \sigma^{JK}\} \psi \bigg] + \dfrac{\hs}{64 \kappa} \bigg\lbrace \left( \dfrac{n-1}{n-2} \right) q_{IJ} \left[ \bar{\psi} \gamma^{I} (E-E^{\dagger}) \psi \right] \left[ \bar{\psi} \gamma^{J} (E-E^{\dagger}) \psi \right]  \notag \\
				&&  + \left( \bar{\psi} \{\gamma^{I}, \sigma^{JK}\} \psi \right) \left( \bar{\psi} \{\gamma_{I}, \sigma_{JK}\} \psi \right) + 3 n_{I} n^{J}  \left( \bar{\psi} \{\gamma^{I}, \sigma^{KL}\} \psi \right) \left( \bar{\psi} \{\gamma_{J}, \sigma_{KL}\} \psi \right)  \bigg\rbrace \notag \\
				\label{const_q_H}
				&&  - \hs \left( 2 \kappa \Lambda + m \bar{\psi} \psi  \right).			
			\end{eqnarray}
		\end{subequations}
	\end{widetext}

	We emphasize that in the formulation~\eqref{S_q}, which comes out after making the symplectomorphism, the field variables are $(\uac{N}, N^a, \tilde{\Pi}^{aI}, \Q_{aI}, \lambda_{IJ}, \psi, \bar{\psi})$, from which $(\uac{N}, N^a, \lambda_{IJ})$ play the role of Lagrange multipliers and $(\tilde{\Pi}^{aI}, \Q_{aI}, \psi, \bar{\psi})$ are the phase-space variables. Note that the phase-space variables $\Q_{aI}$ and $\tilde{\Pi}^{aI}$ transform as a 1-form and as a vector density of weight $+1$ under spatial diffeomorphisms, respectively. However, only $\tilde{\Pi}^{aI}$ transforms as a vector under local Lorentz transformations. The transformation law for $\Q_{aI}$ is a little more complicated, so it does not have a clear geometrical interpretation. Nevertheless, it is worth mentioning that this Hamiltonian formulation also maintains the Lorentz invariance intact and no boundary term is neglected when the definition of $\Q_{aI}$ is made.
	
	The particular cases when the spacetime dimension $n$ is even or odd are similar to those already found at the end of Sec.~\ref{Sec_HA}, and so we do not give further details.
	
	\subsection{Half-densitized fermions}
	The use of half-densitized fermions simplifies the expressions in the Hamiltonian analysis and facilitates the introduction of fermions in the quantization scheme \cite{Thiemann9805, Thiemann9806} (see also Refs.~\cite{Bojowald0809,Romero2106}). Thus, we explore this alternative and define half-densitized fermion fields by
	\begin{subequations}
		\begin{eqnarray}
			\label{hf1}
			\phi &:= &  h^{\frac{1}{4(n-2)}} \psi, \\
			\label{hf2}
			\bar{\phi} &:= & h^{\frac{1}{4(n-2)}} \bar{\psi}. 
		\end{eqnarray}
	\end{subequations}

	Additionally, we rewrite the first term of the action~\eqref{S_Q} as
	\begin{eqnarray}
		&&2 \kappa \tilde{\Pi}^{aI} \dot{Q}_{aI} + \dfrac{1}{2} \h n_{I} \Big( \bar{\psi} \gamma^{I} E \dot{\psi} -  \dot{\bar{\psi}} \gamma^{I} E^{\dagger} \psi \Big) \notag \\
		&&= 2 \kappa \tilde{\Pi}^{aI} \partial_{t} \bigg\lbrace Q_{aI} + \dfrac{1}{8 \kappa (n-2)} \uacc{h}_{ab} \Big[ \tilde{\Pi}^{b}{}_{I} n_{J}  \notag \\
		&&\quad - (n-2) \tilde{\Pi}^{b}{}_{J} n_{I}\Big] \bar{\phi} \gamma^{J} (E - E^{\dagger}) \phi \bigg\rbrace  + \dfrac{1}{2} n_{I} \Big(  \bar{\phi} \gamma^{I} \dot{\phi}  \notag \\
		\label{ss_Psi}
		&& \quad - \dot{\bar{\phi}} \gamma^{I}  \phi \Big) - \dfrac{1}{4(n-2)} \partial_{t} \left[ n_{I} \bar{\phi} \gamma^{I} \left( E - E^{\dagger} \right)  \phi \right]. 
	\end{eqnarray}
	Therefore, it is natural to identify the gravitational variables as
	\begin{eqnarray}
		\label{Psi_def}
		\Psi_{aI} & := & Q_{aI} + \dfrac{1}{8 \kappa (n-2)} \uacc{h}_{ab} \left[ \tilde{\Pi}^{b}{}_{I} n_{J} - (n-2) \tilde{\Pi}^{b}{}_{J} n_{I}\right] \notag \\
		& & \times \bar{\phi} \gamma^{J} (E - E^{\dagger}) \phi.
	\end{eqnarray}
	Note that the boundary term in~\eqref{ss_Psi} is real, so neglecting it does not affect the real character of the symplectic structure. Note also that in the case of the minimal coupling of fermions to gravity, $E=\mathds{1}$, and then $E-E^{\dagger}=0$, so $\Psi_{aI} = Q_{aI}$.
	
	Continuing with the analysis, we use \eqref{hf1}, \eqref{hf2}, and the new variable \eqref{Psi_def} to rewrite the Hamiltonian formulation given in~\eqref{S_Q}. After neglecting the boundary term of \eqref{ss_Psi}, we arrive at
	\begin{eqnarray}
		\label{S_hf}
		S & = & \int_{\mathbb{R} \times \Sigma} dtd^{n-1}x \bigg[  2 \kappa \tilde{\Pi}^{aI} \dot{\Psi}_{aI} + \dfrac{1}{2} n_{I} \Big( \bar{\phi} \gamma^{I} \dot{\phi}  - \dot{\bar{\phi}} \gamma^{I}  \phi \Big) \notag \\ 
		&&  - \lambda_{IJ} \tilde{\mathcal{G}}^{IJ} - 2 N^{a} \tilde{\mathcal{D}}_{a} - \uac{N} \tilde{\tilde{\mathcal{H}}} \bigg],
	\end{eqnarray}
	where the first-class constraints are given by 
	\begin{widetext}
		\begin{subequations} 
			\label{const_hf}
			\begin{eqnarray}
				\label{const_hf_G}
				\tilde{\mathcal{G}}^{IJ} & = & 2 \kappa \tilde{\Pi}^{a[I} \Psi_{a}{}^{J]}
				+ \dfrac{1}{4} n_{K} \bar{\phi} \{\gamma^{K}, \sigma^{IJ}\} \phi, \\
				\label{const_hf_D}
				\tilde{\mathcal{D}}_{a} & = & \kappa \Big( 2 \tilde{\Pi}^{bI} \partial_{[a} \Psi_{b]I} - \Psi_{aI} \partial_{b} \tilde{\Pi}^{bI} \Big) + \frac{1}{4} n_{I}  \left( \bar{\phi} \gamma^{I}  \partial_{a} \phi - \partial_{a} \bar{\phi} \gamma^{I} \phi \right), \\
				\tilde{\tilde{\mathcal{H}}} & = & \kappa \tilde{\Pi}^{aI} \tilde{\Pi}^{bJ} R_{abIJ} + 2 \kappa \tilde{\Pi}^{a[I} \tilde{\Pi}^{|b|J]} \Psi_{aI} \Psi_{bJ} + \dfrac{1}{2} \tilde{\Pi}^{aI} \left( \bar{\phi} \gamma_{I} \nabla_{a} \phi - \overline{\nabla_{a} \phi} \gamma_{I} \phi \right) \notag \\
				&& - \dfrac{1}{2} n_{I} \Psi_{aJ} \tilde{\Pi}^{a}{}_{K} \bar{\phi} \{\gamma^{I}, \sigma^{JK}\} \phi + \dfrac{1}{64 \kappa} \bigg\lbrace \left( \dfrac{n-1}{n-2} \right)  \left[ \bar{\phi} \gamma^{I} (E-E^{\dagger}) \phi \right] \left[ \bar{\phi} \gamma_{I} (E-E^{\dagger}) \phi \right]  \notag \\
				&&  +  \left( \bar{\phi} \{\gamma^{I}, \sigma^{JK}\} \phi \right) \left( \bar{\phi} \{\gamma_{I}, \sigma_{JK}\} \phi \right)  + 3 n_{I} n^{J}  \left( \bar{\phi} \{\gamma^{I}, \sigma^{KL}\} \phi \right) \left( \bar{\phi} \{\gamma_{J}, \sigma_{KL}\} \phi \right)  \bigg\rbrace \notag \\
				\label{const_hf_H}
				&&   - 2 \hs \kappa \Lambda - \h m \bar{\phi} \phi  ,			
			\end{eqnarray}
		\end{subequations}
	with
	\begin{subequations}
		\begin{eqnarray}
			\label{n_phi}
			\nabla_{a} \phi &:=& \partial_{a} \phi - \dfrac{1}{2} \Gamma^{b}{}_{ba} \phi + \dfrac{1}{2} \Gamma_{aIJ} \sigma^{IJ} \phi, \\
			\label{n_bphi}
			\overline{\nabla_{a} \phi} &:=& \partial_{a} \bar{\phi} - \dfrac{1}{2} \Gamma^{b}{}_{ba} \bar{\phi} - \dfrac{1}{2} \Gamma_{aIJ} \bar{\phi} \sigma^{IJ}.
		\end{eqnarray}
	\end{subequations}
    Note that if the coupling is minimal, $E=\mathds{1}$, then the Hamiltonian constraint simplifies more.
	
	The Hamiltonian formulation~\eqref{S_hf} can still be rewritten by factoring out a term involving the Gauss constraint in the Hamiltonian constraint. The term involving the Gauss constraint can be explicitly displayed by using the constraint \eqref{const_hf_G}, which allows us to rewrite \eqref{const_hf_H} as
		\begin{eqnarray}
			\tilde{\tilde{\mathcal{H}}} & = & \kappa \tilde{\Pi}^{aI} \tilde{\Pi}^{bJ} R_{abIJ} + 2 \kappa \tilde{\Pi}^{a[I} \tilde{\Pi}^{|b|J]} \Psi_{aI} \Psi_{bJ} + \dfrac{1}{2} \tilde{\Pi}^{aI} \left( \bar{\phi} \gamma_{I} \nabla_{a} \phi - \overline{\nabla_{a} \phi} \gamma_{I} \phi \right) \notag \\
			&&  + \dfrac{1}{64 \kappa} \bigg\lbrace \left( \dfrac{n-1}{n-2} \right)  \left[ \bar{\phi} \gamma^{I} (E-E^{\dagger}) \phi \right] \left[ \bar{\phi} \gamma_{I} (E-E^{\dagger}) \phi \right]  + \left( \bar{\phi} \{\gamma^{I}, \sigma^{JK}\} \phi \right) \left( \bar{\phi} \{\gamma_{I}, \sigma_{JK}\} \phi \right) \notag \\
			&&    -  n_{I} n^{J}  \left( \bar{\phi} \{\gamma^{I}, \sigma^{KL}\} \phi \right) \left( \bar{\phi} \{\gamma_{J}, \sigma_{KL}\} \phi \right)  \bigg\rbrace - 2 \hs \kappa \Lambda - \h m \bar{\phi} \phi + \dfrac{1}{4 \kappa} \tilde{\mathcal{G}}_{IJ} n_{K} \bar{\phi} \{\gamma^{K}, \sigma^{IJ}\} \phi. 
			\label{const_hf_H_2}
		\end{eqnarray}
	
	Therefore, factoring out the term involving the Gauss constraint, which requires us to redefine the Lagrange multiplier $\mu_{IJ} := \lambda_{IJ} + (1/4) \kappa^{-1} \uac{N} n^{K} \bar{\phi} \{\gamma_{K}, \sigma_{IJ}\} \phi$, we get the Hamiltonian formulation with half-densitized fermion fields
	\begin{eqnarray}
		\label{S_hf_2}
		S & = & \int_{\mathbb{R} \times \Sigma} dtd^{n-1}x \bigg[  2 \kappa \tilde{\Pi}^{aI} \dot{\Psi}_{aI} + \dfrac{1}{2} n_{I} \Big( \bar{\phi} \gamma^{I} \dot{\phi}  - \dot{\bar{\phi}} \gamma^{I}  \phi \Big)  - \mu_{IJ} \tilde{\mathcal{G}}^{IJ} - 2 N^{a} \tilde{\mathcal{D}}_{a} - \uac{N} \tilde{\tilde{\mathcal{C}}} \bigg],
	\end{eqnarray}
	where the Gauss and diffeomorphism constraints are given by \eqref{const_hf_G} and \eqref{const_hf_D}, respectively, and the Hamiltonian constraint is 	
		\begin{eqnarray}
			\tilde{\tilde{\mathcal{C}}} & := & \kappa \tilde{\Pi}^{aI} \tilde{\Pi}^{bJ} R_{abIJ} + 2 \kappa \tilde{\Pi}^{a[I} \tilde{\Pi}^{|b|J]} \Psi_{aI} \Psi_{bJ} + \dfrac{1}{2} \tilde{\Pi}^{aI} \left( \bar{\phi} \gamma_{I} \nabla_{a} \phi - \overline{\nabla_{a} \phi} \gamma_{I} \phi \right) \notag \\
			&&  + \dfrac{1}{64 \kappa} \bigg\lbrace \left( \dfrac{n-1}{n-2} \right)  \left[ \bar{\phi} \gamma^{I} (E-E^{\dagger}) \phi \right] \left[ \bar{\phi} \gamma_{I} (E-E^{\dagger}) \phi \right]  +  \left( \bar{\phi} \{\gamma^{I}, \sigma^{JK}\} \phi \right) \left( \bar{\phi} \{\gamma_{I}, \sigma_{JK}\} \phi \right) \notag \\
			\label{const_hf_C}
			&&    -  n_{I} n^{J}  \left( \bar{\phi} \{\gamma^{I}, \sigma^{KL}\} \phi \right) \left( \bar{\phi} \{\gamma_{J}, \sigma_{KL}\} \phi \right) \bigg\rbrace - 2 \hs \kappa \Lambda - \h m \bar{\phi} \phi .	
		\end{eqnarray}
	\end{widetext}
	
	We stress that in the formulation~\eqref{S_hf_2}, which comes out from using of half-densitized fermions, the field variables are $(\uac{N}, N^a, \tilde{\Pi}^{aI}, \Psi_{aI}, \mu_{IJ}, \phi, \bar{\phi})$, from which $(\uac{N}, N^a, \mu_{IJ})$ play the role of Lagrange multipliers and $(\tilde{\Pi}^{aI}, \Psi_{aI}, \phi, \bar{\phi})$ are the phase-space variables. Note that the phase-space variables $(\Psi_{aI}, \tilde{\Pi}^{aI})$ transform as Lorentz vectors under local $SO(n-1,1)$ transformations. Furthermore, this formulation generalizes in two aspects the one presented in Ref.~\cite{Bodendorfer1301d}, where authors consider the time gauge from the very beginning of their analysis and also the Hamiltonian formulation is restricted to the case of the minimal coupling of fermions, $E=\mathds{1}$. We also explore the time gauge, but in Sec.~\ref{Sec_tg}. Regarding the coupling, note that the matrix $E$ appears only in the scalar constraint, in some of the terms involving the quartic fermion interaction. In the case where the spacetime dimension is even, the scalar constraint is
	\begin{widetext}
		\begin{eqnarray}
			\tilde{\tilde{\mathcal{C}}} & = & \kappa \tilde{\Pi}^{aI} \tilde{\Pi}^{bJ} R_{abIJ} + 2 \kappa \tilde{\Pi}^{a[I} \tilde{\Pi}^{|b|J]} \Psi_{aI} \Psi_{bJ} + \dfrac{1}{2} \tilde{\Pi}^{aI} \left( \bar{\phi} \gamma_{I} \nabla_{a} \phi - \overline{\nabla_{a} \phi} \gamma_{I} \phi \right) \notag \\
			&&  + \dfrac{1}{64 \kappa} \bigg[ 4 \left( \dfrac{n-1}{n-2} \right)  \left( \theta^{2} \tilde{V}_{I} \tilde{V}^{I} + \xi^{2} \tilde{A}_{I} \tilde{A}^{I} + 2 \theta \xi \tilde{V}_{I} \tilde{A}^{I} \right)   +   \left( \bar{\phi} \{\gamma^{I}, \sigma^{JK}\} \phi \right) \left( \bar{\phi} \{\gamma_{I}, \sigma_{JK}\} \phi \right)  \notag \\
			\label{const_hf_H_e}
			&& - n_{I} n^{J}  \left( \bar{\phi} \{\gamma^{I}, \sigma^{KL}\} \phi \right) \left( \bar{\phi} \{\gamma_{J}, \sigma_{KL}\} \phi \right) \bigg]  - 2 \hs \kappa \Lambda - \h m \bar{\phi} \phi .
		\end{eqnarray}

		On the other hand, when $n$ is odd, the scalar constraint becomes
		\begin{eqnarray}
			\tilde{\tilde{\mathcal{C}}} & = & \kappa \tilde{\Pi}^{aI} \tilde{\Pi}^{bJ} R_{abIJ} + 2 \kappa \tilde{\Pi}^{a[I} \tilde{\Pi}^{|b|J]} \Psi_{aI} \Psi_{bJ} + \dfrac{1}{2} \tilde{\Pi}^{aI} \left( \bar{\phi} \gamma_{I} \nabla_{a} \phi - \overline{\nabla_{a} \phi} \gamma_{I} \phi \right) \notag \\
			&& + \dfrac{1}{64 \kappa} \bigg[ 4 \theta^{2} \left( \dfrac{n-1}{n-2} \right) \tilde{V}_{I} \tilde{V}^{I}  + \left( \bar{\phi} \{\gamma^{I}, \sigma^{JK}\} \phi \right) \left( \bar{\phi} \{\gamma_{I}, \sigma_{JK}\} \phi \right) \notag \\
			\label{const_hf_H_o}
			&&    - n_{I} n^{J}  \left( \bar{\phi} \{\gamma^{I}, \sigma^{KL}\} \phi \right) \left( \bar{\phi} \{\gamma_{J}, \sigma_{KL}\} \phi \right) \bigg] - 2 \hs \kappa \Lambda - \h m \bar{\phi} \phi .			
		\end{eqnarray}
	\end{widetext}
	Regardless of the spacetime dimension, the real densitized fermion currents are defined by
	\begin{subequations}
		\begin{eqnarray}
			\label{Vcd_def}
			\tilde{V}^{I} &:=& \mathrm{i} \bar{\phi} \gamma^{I} \phi, \\
			\label{Acd_def}
			\tilde{A}^{I} &:=& \mathrm{i} \bar{\phi} \Gamma \gamma^{I} \phi.
		\end{eqnarray}
	\end{subequations}
	
	\section{Time gauge}
	\label{Sec_tg}
	In Secs.~\ref{Sec_HA} and~\ref{Sec_2more} we have presented several Hamiltonian formulations involving manifestly Lorentz-covariant phase-space variables. To make contact with other Hamiltonian formulations, we impose the time gauge, fixing the freedom to perform boosts, and reducing the gauge symmetry to (the double cover of) $SO(n-1)$.
	
	The time gauge is given by  
	\begin{equation}
		\label{GC}
		\tilde{\Pi}^{a0} = 0,
	\end{equation}
	which is equivalent to $n_{i}=0$ [see Eq.~\eqref{n} and Appendix~\ref{ApendiceA}] as long as $\det(\tilde{\Pi}^{ai})\neq 0$, which is assumed throughout this section. The condition \eqref{GC} together with $\tilde{\mathcal{G}}^{i0}$ are second-class constraints because 
	\begin{equation}
		\{\tilde{\Pi}^{a0}(t,x), \tilde{\mathcal{G}}^{i0}(t,y)\} = \dfrac{1}{2}\tilde{\Pi}^{ai} \delta^{n-1}(x,y).
	\end{equation}
	Therefore, enforcing the condition \eqref{GC} requires to solve the constraints $\tilde{\mathcal{G}}^{i0}=0$, which depends on each of the Hamiltonian formulations presented previously. Thus, let us analyze each of them. Moreover, from the definition \eqref{CDP}, note that \eqref{GC} also implies $\Gamma_{a0i}=0$ and $\Gamma_{aij}$ becomes the spin connection compatible with the densitized frame $\tilde{\Pi}^{ai}$. 
	
	\subsection{Time gauge in the Hamiltonian formulation involving $Q_{aI}$}
	
	We impose the time gauge in the Hamiltonian formulation given by the action~\eqref{S_Q} derived in Sec.~\ref{Sec_HA}. Thus, from \eqref{const_Q_G}, the solution of $\tilde{\mathcal{G}}^{i0}=0$ is 
	\begin{eqnarray}
		\label{Q0}
		Q_{a0} &=& \dfrac{n_{0}}{8\kappa} \h \uac{\Pi}_{ai} \bar{\psi} \gamma^{i} (E - E^{\dagger}) \psi,
	\end{eqnarray}
	where $n_{0} = \mbox{sgn}[\det(\tilde{\Pi}^{ai})]$ is the only nonzero component of $n_{I}$ left [see Eq.~\eqref{n}] with $\uac{\Pi}_{ai}$ being the inverse of $\tilde{\Pi}^{ai}$. Using \eqref{GC} and \eqref{Q0}, the action~\eqref{S_Q} acquires the form
	\begin{eqnarray}
		\label{S_Q_tg}
		S & = & \int_{\mathbb{R} \times \Sigma} dtd^{n-1}x \bigg[  2 \kappa \tilde{\Pi}^{ai} \dot{Q}_{ai} + \dfrac{1}{2} \h n_{0} \Big( \bar{\psi} \gamma^{0} E \dot{\psi}  \notag \\
		&&-  \dot{\bar{\psi}} \gamma^{0} E^{\dagger} \psi \Big) - \lambda_{ij} \tilde{\mathcal{G}}^{ij} - 2 N^{a} \tilde{\mathcal{D}}_{a} - \uac{N} \tilde{\tilde{\mathcal{H}}} \bigg],
	\end{eqnarray}
	where the first-class constraints are given by
	\begin{widetext}
		\begin{subequations} 
			\label{const_Q_tg}
			\begin{eqnarray}
				\label{const_Q_G_tg}
				\tilde{\mathcal{G}}^{ij} & = & 2 \kappa \tilde{\Pi}^{a[i} Q_{a}{}^{j]}
				+ \dfrac{n_{0}}{2} \h  \bar{\psi} \gamma^{0} \sigma^{ij} \psi , \\
				\tilde{\mathcal{D}}_{a} & = & \kappa \Big( 2 \tilde{\Pi}^{bi} \partial_{[a} Q_{b]i} - Q_{ai} \partial_{b} \tilde{\Pi}^{bi} \Big) + \frac{n_{0}}{4} \h   \left( \bar{\psi} \gamma^{0} E  \partial_{a} \psi - \partial_{a} \bar{\psi} \gamma^{0} E^{\dagger} \psi \right), \\
				\tilde{\tilde{\mathcal{H}}} & = & \kappa \tilde{\Pi}^{ai} \tilde{\Pi}^{bj} R_{abij} + 2 \kappa \tilde{\Pi}^{a[i} \tilde{\Pi}^{|b|j]} Q_{ai} Q_{bj} + \dfrac{1}{2} \h \tilde{\Pi}^{ai} \left( \bar{\psi} \gamma_{i} \nabla_{a} \psi - \overline{\nabla_{a} \psi} \gamma_{i} \psi \right) \notag \\
				&& + n_{0} \h  Q_{ai}  \left[ \dfrac{1}{4}\tilde{\Pi}^{ai}  \bar{\psi} \gamma^{0}(E - E^{\dagger}) \psi -  \tilde{\Pi}^{a}{}_{j} \bar{\psi} \gamma^{0} \sigma^{ij} \psi \right] \notag \\
				&& + \dfrac{\hs}{64 \kappa} \bigg\lbrace \left( \dfrac{n-1}{n-2} \right) \left[ \bar{\psi} \gamma^{i} (E-E^{\dagger}) \psi \right] \left[ \bar{\psi} \gamma_{i} (E-E^{\dagger}) \psi \right]  + \left[ \bar{\psi} \{\gamma^{i}, \sigma^{jk}\} \psi \right] \left[ \bar{\psi} \{\gamma_{i}, \sigma_{jk}\} \psi \right]  \bigg\rbrace\notag \\
				\label{const_Q_H_tg}
				&&   - \hs \left( 2 \kappa \Lambda + m \bar{\psi} \psi  \right).			
			\end{eqnarray}
		\end{subequations}

	This Hamiltonian formulation is also the one obtained imposing the time gauge in the action~\eqref{S_q} because in the time gauge $Q_{ai} = \Q_{ai}$ [see Eq.~\eqref{Q2q}].

	\subsection{Time gauge in the Hamiltonian formulation involving  half-densitized fermions}
	
	We now impose the time gauge in the Hamiltonian formulation involving half-densitized fermions. We use the formulation encompassed by the action~\eqref{S_hf_2} and the constraints \eqref{const_hf_G}, \eqref{const_hf_D}, and \eqref{const_hf_C}. Using \eqref{const_hf_G} and solving $\tilde{\mathcal{G}}^{i0}=0$, we get
	\begin{equation}
		\Psi_{a0} = 0.
	\end{equation}
	Therefore, the ensuing formulation is described by the action
	\begin{eqnarray}
		\label{S_hf_tg}
		S & = & \int_{\mathbb{R} \times \Sigma} dtd^{n-1}x \bigg[  2 \kappa \tilde{\Pi}^{ai} \dot{\Psi}_{ai} + \dfrac{1}{2} n_{0} \Big( \bar{\phi} \gamma^{0} \dot{\phi}  - \dot{\bar{\phi}} \gamma^{0}  \phi \Big)   - \mu_{ij} \tilde{\mathcal{G}}^{ij} - 2 N^{a} \tilde{\mathcal{D}}_{a} - \uac{N} \tilde{\tilde{\mathcal{H}}} \bigg],
	\end{eqnarray}
	and the first-class constraints are given by
		\begin{subequations} 
			\label{const_hf_tg}
			\begin{eqnarray}
				\label{const_hf_G_tg}
				\tilde{\mathcal{G}}^{ij} & = & 2 \kappa \tilde{\Pi}^{a[i} \Psi_{a}{}^{j]}
				+ \dfrac{n_{0}}{2} \bar{\phi} \gamma^{0} \sigma^{ij} \phi, \\
				\tilde{\mathcal{D}}_{a} & = & \kappa \Big( 2 \tilde{\Pi}^{bi} \partial_{[a} \Psi_{b]i} - \Psi_{ai} \partial_{b} \tilde{\Pi}^{bi} \Big) + \frac{n_{0}}{4}   \left( \bar{\phi} \gamma^{0}  \partial_{a} \phi - \partial_{a} \bar{\phi} \gamma^{0} \phi \right), \\
				\tilde{\tilde{\mathcal{H}}} & = & \kappa \tilde{\Pi}^{ai} \tilde{\Pi}^{bj} R_{abij} + 2 \kappa \tilde{\Pi}^{a[i} \tilde{\Pi}^{|b|j]} \Psi_{ai} \Psi_{bj} + \dfrac{1}{2} \tilde{\Pi}^{ai} \left( \bar{\phi} \gamma_{i} \nabla_{a} \phi - \overline{\nabla_{a} \phi} \gamma_{i} \phi \right) \notag \\
				&&  + \dfrac{1}{64 \kappa} \bigg\lbrace \left( \dfrac{n-1}{n-2} \right)  \left[ \bar{\phi} \gamma^{I} (E-E^{\dagger}) \phi \right] \left[ \bar{\phi} \gamma_{I} (E-E^{\dagger}) \phi \right] - 16 \left(\bar{\phi} \gamma^{0} \sigma^{ij} \phi \right) \left(\bar{\phi} \gamma^{0} \sigma_{ij} \phi \right)  \notag \\
				\label{const_hf_H_tg}
				&&  +  \left(\bar{\phi} \{\gamma^{i}, \sigma^{jk}\} \phi \right) \left( \bar{\phi} \{\gamma_{i}, \sigma_{jk}\} \phi \right) \bigg\rbrace  - 2 \hs \kappa \Lambda - \h m \bar{\phi} \phi  .	
			\end{eqnarray}
		\end{subequations}
	\end{widetext}

	Note that in the case of minimal coupling, $E=\mathds{1}$, this formulation becomes the one derived in Ref.~\cite{Bodendorfer1301d}. This is other way of seeing that the formulation presented in \eqref{S_hf_2} with the constraints \eqref{const_hf_G}, \eqref{const_hf_D}, and \eqref{const_hf_C} [restricted to the case $E=\mathds{1}$] is indeed a manifestly Lorentz-invariant generalization of the one of Ref.~\cite{Bodendorfer1301d}.
	
	\section{Concluding remarks}\label{Sec_concl}
	In this paper we have carried out the Lagrangian analysis of a fermion field minimally and nonminimally coupled to the Palatini action in $n$ dimensions. A remarkable fact is that in a spacetime of dimension four, there exists a first-order Lagrangian action with a particular nonminimal coupling of the fermion field to the Palatini action that is equivalent to the Einstein-Dirac action plus a boundary term. This result is analogous to what happens for the specific nonminimal coupling of the fermion field to the Holst action studied in Ref.~\cite{Romero2106}, where the interaction term of the resulting second-order action also vanishes. Nevertheless, the Lagrangian action considered in this paper and the one of Ref.~\cite{Romero2106} have different coupling matrix (see Sec.~\ref{Sec_Lagran} of this paper).
	
	Regarding the Hamiltonian analysis, it is important to emphasize that all the Hamiltonian formulations presented in Secs.~\ref{Sec_HA} and~\ref{Sec_2more} of this paper involve manifestly Lorentz-covariant phase-space variables and that their corresponding symplectic structures are both Lorentz-invariant and real. Additionally, the local Lorentz symmetry displays itself in the first-class constraints through the various coupling terms present there, which strongly contrasts with the form of the symplectic structure and the form of the first-class constraints of the Hamiltonian formulations when the time gauge is imposed, which are presented in Sec.~\ref{Sec_tg}. 
	
	It is also important to mention that the Hamiltonian formulation~\eqref{S_hf_tg} reduces to the one presented in Ref.~\cite{Bodendorfer1301d} when the coupling of the fermion field is minimal ($E=\mathds{1}$). Consequently, the Hamiltonian formulation~\eqref{S_hf_2} ---from which~\eqref{S_hf_tg} comes from imposing the time gauge---is a generalization in two aspects of the one reported in Ref.~\cite{Bodendorfer1301d}.
	
	Finally, it is also important to remark that additional manifestly Lorentz-covariant Hamiltonian formulations can be found by making symplectomorphisms along the lines of the ones considered in Refs.~\cite{Montesinos2001,Montesinos2004a,Romero2106}.

	\acknowledgments
	We thank Mariano Celada for his valuable comments. This work was partially supported by Consejo Nacional de Ciencia y Tecnología (CONACyT), M\'{e}xico, Grant No. A1-S-7701.

	\appendix

	\section{Conventions and notation}\label{ApendiceA}
	
	\subsection{General relativity}
	\label{Ap_gc}
	We label the points on $M$ with local coordinates $\{x^{\mu}\} = \{ t:=x^{0}, x^{a}\}$, where lower case latin indices $a,b,c,\ldots$ take on the values $1,\ldots, n-1$. In every point $x\in M$, we have the cotangent space, where we have an orthonormal frame of 1-forms $e^{I}$, i.e.,  $g = \eta_{IJ} e^I \otimes e^J$ where $g$ is the metric tensor and $(\eta_{IJ}) = \mbox{diag}(-1,1,\ldots, 1)$ is the Minkowski metric. Thus, the indices $I, J, K, \ldots$ that take on the values $0,1,\dots, n-1$ are $SO(n-1,1)$ valued and are lowered and raised with $\eta_{IJ}$. Similarly, the connection $\omega^{I}{}_{J}$ is compatible with $\eta_{IJ}$, $D\eta_{IJ} := d\eta_{IJ} - \omega^{K}{}_{I} \eta_{KJ} - \omega^{K}{}_{J} \eta_{IK} =0$, and so $\omega_{IJ} = - \omega_{JI}$. In the first-order formalism of general relativity, the orthonormal frame of 1-forms (vielbein) $e^{I}$ and the connection $\omega^{I}{}_{J}$ are the fundamental independent variables of the theory. The symbol ``$\star$'' is the Hodge dual defined by
	\begin{equation}
		\label{Hd}
		\star \left( e_{I_{1}} \wedge \cdots \wedge e_{I_{k}} \right) := \dfrac{1}{(n-k)!} \epsilon_{I_{1} \ldots I_{k} I_{k+1} \ldots I_{n}} e^{I_{k+1}} \wedge \cdots \wedge e^{I_{n}},
	\end{equation}
	where the totally antisymmetric Lorentz tensor $\epsilon_{I_{1} \ldots I_{n}}$ is such that $\epsilon_{01\ldots (n-1)} =1$. Symmetrization and antisymmetrization of Lorentz indices are denoted, respectively, by
	\begin{eqnarray}
		A_{(IJ)} &=& \frac12 \left ( A_{IJ} + A_{JI} \right ), \nonumber\\
		A_{[IJ]} &=& \frac12 \left ( A_{IJ} - A_{JI} \right ).
	\end{eqnarray}
	Similarly, symmetrization and antisymmetrization of space indices are denoted, respectively, by
	\begin{eqnarray}
		A_{(ab)} &=& \frac12 \left ( A_{ab} + A_{ba} \right ), \nonumber\\
		A_{[ab]} &=& \frac12 \left ( A_{ab} - A_{ba} \right ).
	\end{eqnarray}
	Moreover, indices inside the vertical bars $\mid \mid$, as in $A_{(a \mid IJ cd \mid b)}$, are not symmetrized. Also, the indices inside $\mid\mid$ in $A_{[a \mid IJ cd \mid b]}$ are not antisymmetrized. 
	
	Furthermore, $\etatu_{a_{1}\ldots a_{n-1}}$ is the totally antisymmetric tensor density of weight $-1$ on $\Sigma$ such that $\etatu_{1 23\ldots (n-1)} =1$. Similarly, $\etat^{a_{1} \ldots a_{n-1}}$ is the totally antisymmetric tensor density of weight $+1$ such that $\etat^{12\ldots (n-1)} =1$. Also, two tildes above a tensor means it is of weight $+2$ and two tildes below a tensor means it is of weight $-2$. However, to avoid a cumbersome notation, sometimes the weight is not indicated with tildes, but the weight is explicitly mentioned in the text. 
	
	When the time gauge is imposed (Sec.~\ref{Sec_tg}), the indices $I,J,K, \ldots$ split into ``0'' and the spatial internal indices $i,j,k, \ldots$ that take on the values $1, \ldots,n-1$.
	
	\subsection{Fermion field}
	\label{Ap_fc}
	The fermion field $\psi$ is Grassmann-valued and $\bar{\psi}:= \mathrm{i} \psi^{\dagger} \gamma^{0}$ is its Dirac conjugate, where 
	$\mathrm{i}$ is the imaginary unit. The Dirac matrices $\{\gamma^{I}\}= \{ \gamma^0, \gamma^1, \ldots, \gamma^{n-1}\}$ satisfy the Clifford algebra 
	\begin{equation}
		\label{Cliff_alg}
		\{\gamma^{I}, \gamma^{J}\} := \gamma^{I}\gamma^{J} + \gamma^{J} \gamma^{I} = 2 \eta^{IJ} \mathds{1},
	\end{equation}
	where $\mathds{1}$ is the $2^{[n/2]}\times 2^{[n/2]}$ identity matrix ($[n/2]$ denotes the integer part of $n/2$). Note also that $(\gamma^{I})^{\dagger} := \gamma^{0} \gamma^{I} \gamma^{0}$.
	
	Furthermore, the Lorentz generators in the spin representation are the $n(n-1)/2$ quantities $\sigma^{IJ} = - \sigma^{JI} := (1/4) [\gamma^{I}, \gamma^{J}]$. Thus, the covariant derivatives of $\psi$ and $\bar{\psi}$ with respect $\omega^{I}{}_{J}$ are
	\begin{subequations}
		\begin{eqnarray}
			\label{Dp}
			D \psi &:= & d \psi + \frac{1}{2} \omega_{IJ} \sigma^{IJ} \psi, \\ 
			\label{Dbp}
			\overline{D \psi} &:= &  d \bar{\psi} - \frac{1}{2} \omega_{IJ} \bar{\psi} \sigma^{IJ}.
		\end{eqnarray}
	\end{subequations}
	
	From the definition of $\sigma^{IJ}$ and \eqref{Cliff_alg}, we have
	\begin{equation}
		\label{anti_comm}
		[\gamma^{I}, \sigma^{JK}] = \gamma^{I} \sigma^{JK} - \sigma^{JK} \gamma^{I} = 2 \eta^{I[J} \gamma^{K]}.
	\end{equation}
	
	Similarly, the identity
	\begin{equation}
		\label{g3}
		\gamma^{I} \gamma^{J} \gamma^{K} = \eta^{JK} \gamma^{I} - \eta^{IK} \gamma^{J} + \eta^{IJ} \gamma^{K} + \{\gamma^{I}, \sigma^{JK}\}
	\end{equation}
	holds. Furthermore, we have
	\begin{equation}
		\label{id_anticomm}
		\{\gamma^{I}, \sigma^{JK}\} = C_{n} \epsilon^{IJKL_{1}\cdots L_{n-3}} \Gamma \gamma_{L_{1}} \cdots \gamma_{L_{n-3}},
	\end{equation}
	where 
	\begin{equation}
		\label{C_n}
		C_{n} := - \dfrac{(-1)^{n-3}}{(n-3)!} \mathrm{i}^{\frac{n(n-1) -2}{2}},
	\end{equation}
	and $\Gamma$ is the chirality matrix:
	\begin{equation}
		\label{CM}
		\Gamma := \mathrm{i}^{\frac{n(n-1)-2}{2} } \gamma^{0} \cdots \gamma^{n-1}.
	\end{equation}
	Defined as such, the chirality matrix satisfies $\Gamma^{\dagger} = \Gamma$ and $\Gamma^{2} = \mathds{1}$. Also, when $n$ is even, we have $\gamma^{I} \Gamma = - \Gamma \gamma^{I}$. However, if $n$ is odd, $\gamma^{I} \Gamma = \Gamma \gamma^{I}$. Thus, by Schur's lemma, if $n$ is odd, $\Gamma$ is proportional to the identity matrix $\mathds{1}$. Therefore, we only consider the chirality matrix in the cases where the spacetime dimension $n$ is even. 
	
	\section{Fermion field in Minkowski spacetime}\label{Ap_Mink}
	When gravity and the cosmological constant are turned off, the spacetime $M$ becomes the Minkowski spacetime $\mathcal{M}^{n}$. In a Minkowski spacetime, the fermion action~\eqref{S_F} acquires the form 
	\begin{eqnarray}
		\label{S_F_flat}
		S_{F}[\psi, \bar{\psi}] &=& \int_{\mathcal{M}^{n}} d^{n}x \bigg[ \frac{1}{2}  \left(  \bar{\psi} \gamma^{\mu} E \partial_{\mu} \psi -  \partial_{\mu}\bar{\psi} \gamma^{\mu} E^{\dagger} \psi \right) \notag \\
		&& - m \bar{\psi} \psi \bigg],
	\end{eqnarray}
	where $x^{\mu}$ are Minkowski coordinates and $\gamma^{\mu}$ are the usual $\gamma$ matrices defined in inertial frames. Varying this action results in the equations of motion
	\begin{subequations}
		\begin{eqnarray}
			\dfrac{1}{2} \gamma^{\mu} ( E + E^{\dagger} ) \partial_{\mu} \psi  - m \psi = 0, \\
			- \dfrac{1}{2} \partial_{\mu} \bar{\psi} \gamma^{\mu} ( E + E^{\dagger} )   - m \bar{\psi} = 0.
		\end{eqnarray}
	\end{subequations}
	Therefore, the action~\eqref{S_F_flat} correctly yields the Dirac equation only if $E + E^{\dagger} = 2 \mathds{1}$. This is the reason behind the definition of $E$ given in \eqref{E}. A more general coupling matrix $E = (1+ \mathrm{i} \theta) \mathds{1} - (\lambda + \mathrm{i} \xi ) \Gamma$, with $\lambda \in \mathbb{R}$, does not give the Dirac equation, not even modifying the mass term $m \bar{\psi} \psi$~\cite{Kazmierczak0903}. 
	
	\section{Maps and transformations}\label{Ap_maps}
	
	In Sec.~\ref{Sec_HA}, it is introduced the map $(N, N^{a}, \tilde{\Pi}^{aI})\mapsto (e_{\mu}{}^{I})$ given by \eqref{ep1} and \eqref{ep2}. It is a one-to-one map, whose inverse is given by
	\begin{subequations}
		\begin{eqnarray}
			N &=& - n_{I} e_{0}{}^{I}, \label{inverse_map}\\
			N^{a} &=& q^{ab} e_{0}{}^{I} e_{bI},\\
			\tilde{\Pi}^{aI} &=& \sqrt{q} q^{ab} e_{b}{}^{I},
		\end{eqnarray}
	\end{subequations}
	where $q_{ab}:= e_{a}{}^{I}e_{bI}$ is the spatial metric on $\Sigma$, $q^{ab}$ is its inverse, and $q=\det(q_{ab})$ has weight $2$. Note that in the right-hand side of~\eqref{inverse_map} $n_{I}$ must be understood as 
	\begin{equation}
		\label{nI_e}
		n_{I}= \frac{1}{(n-1)!\sqrt{q}}\epsilon_{I J_1 \ldots J_{n-1}}\tilde{\eta}^{a_1 \ldots a_{n-1}} e_{a_{1}}{}^{J_{1}} \cdots e_{a_{n-1}}{}^{J_{n-1}},
	\end{equation}
	i.e., $n_I$ is a function of $e_{a}{}^{I}$.
	
	On the other hand, \eqref{wQ} defines the map $(Q_{aI}, \uacc{u}_{abc}) \mapsto \omega_{a}{}^{I}{}_{J}$, where $M_{a}{}^{b}{}_{IJK}$ and $\tilde{\tilde{N}}_a{}^{bcd}{}_{IJ}$ are given by
	\begin{eqnarray}
		\label{M}
		M_{a}{}^{b}{}_{I J K} &:=& - \frac{2}{(n-2)}\Big{[} (n-2)\delta^{b}_{a}n_{[I}\eta_{J] K} \notag \\ 
		&&+  \uac{\uac{h}}_{a c}\tilde{\Pi}^{c}{}_{[I}\tilde{\Pi}^{b}{}_{J]}n_{K} \Big{]}, \\
		\label{N}
		\tilde{\tilde{N}}_a{}^{bcd}{}_{IJ}&:=&\bigg(\delta_a^b\delta_e^{[c}\delta_f^{d]} -\frac{2}{n-2}\uac{\uac{h}}_{ae}\tilde{\tilde{h}}^{b[c}\delta_f^{d]} \bigg)\tilde{\Pi}^{e}{}_{[I}\tilde{\Pi}^{f}{}_{J]}. \notag \\
	\end{eqnarray}
	Moreover, defining the object $\uac{\uac{U}}_{abc}{}^{dIJ}= - \uac{\uac{U}}_{acb}{}^{dIJ}= - \uac{\uac{U}}_{abc}{}^{dJI}$ as
	\begin{equation}
		\label{U}
		\uac{\uac{U}}_{abc}{}^{dIJ}  := \left( \delta^{d}_{a}\uac{\uac{h}}_{e[b}\uac{\uac{h}}_{c]f} - \frac{2}{n-2}\uac{\uac{h}}_{a [b}\uac{\uac{h}}_{c] f}\delta^{d}_{e} \right)\tilde{\Pi}^{e [I}\tilde{\Pi}^{|f|J]},
	\end{equation}
	we complete the map $(\omega_{a}{}^{I}{}_{J}) \mapsto (Q_{aI}, \uacc{u}_{abc})$, which is given by \eqref{Q_def} and 
	\begin{eqnarray}
		\label{u_fields}
		\uac{\uac{u}}_{abc} = \uac{\uac{U}}_{abc}{}^{dIJ} \left( \omega_{dIJ} - \Gamma_{dIJ} \right).
	\end{eqnarray}
	Therefore, the map between  $(\omega_{a}{}^{I}{}_{J})$ and $(Q_{aI}, \uacc{u}_{abc})$ is invertible. Furthermore, the objects $M_{a}{}^{b}{}_{IJK}$, $\tilde{\tilde{N}}_a{}^{bcd}{}_{IJ}$, $\uac{\uac{U}}_{abc}{}^{dIJ}$, and $W_{a}{}^{b}{}_{IJK}$ satisfy the relations
	\begin{subequations}
		\begin{eqnarray} 
			\label{WM}
			W_{a}{}^{c I KL}M_{c}{}^{b}{}_{KLJ} &=& \delta^{b}_{a}\delta^{I}_{J}, \\
			\label{UN}
			\uac{\uac{U}}_{cde}{}^{gIJ}  \tilde{\tilde{N}}_g{}^{fab}{}_{IJ} &=&\delta^f_c \delta^{[a}_{d}\delta^{b]}_{e} - \frac{1}{n-2} \Big ( \uac{\uac{h}}_{cd} \tilde{\tilde{h}}^{f[a} \delta^{b]}_e \notag \\
			&& - \uac{\uac{h}}_{ce} \tilde{\tilde{h}}^{f[a} \delta^{b]}_d \Big), \\
			\label{WN}
			W_{a}{}^{f}{}_{IJK}\tilde{\tilde{N}}_f{}^{bcdJK} &=& 0, \\
			\label{UM}
			\uac{\uac{U}}_{abc}{}^{dIJ}M_{d}{}^{e}{}_{IJK} &=& 0.
		\end{eqnarray}
	\end{subequations}
	The presence of the second term in the right-hand side of~\eqref{UN} is a consequence of the conditions $\uac{\uac{h}}_{b c}\tilde{\tilde{N}}_a{}^{bcd}{}_{IJ}=0$ and $\tilde{\tilde{h}}^{ab} \uac{\uac{U}}_{abc}{}^{dIJ}=0$ imposed such that $\uacc{u}_{abc}$ has the correct number of independent fields. Furthermore, we have the completeness relation
	\begin{eqnarray}
		M_{a}{}^{c}{}_{IJM} W_{c}{}^{bMKL} + \tilde{\tilde{N}}_{a}{}^{cdf}{}_{IJ}\uac{\uac{U}}_{cdf}{}^{bKL}  &=& \delta^{b}_{a} \delta^K_{[I} \delta^L_{J]}. \quad \quad
	\end{eqnarray}
	\phantom{.}
	\section{Hamiltonian formulations when $n=3$}
	\label{Ap_3d}
	The parametrization of the vielbein $e^{I}$ and the connection $\omega^{I}{}_{J}$ is given in~\eqref{ep1},~\eqref{ep2},~\eqref{wQ},~and \eqref{lQ} of Sec.~\ref{Sec_HA}. However, when $n=3$ things become easy because there are no auxiliary fields $\uacc{u}_{abc}$, and the parametrization of $\omega_{a}{}^{I}{}_{J}$ is just~\cite{Montesinos2001}
	\begin{equation}\label{w3d}
		\omega_{aIJ} = M_{a}{}^{b}{}_{IJK} Q_{b}{}^{K} + \Gamma_{aIJ}, 
	\end{equation}
	where $M_{a}{}^{b}{}_{IJK}$ is given in~\eqref{M} and $\Gamma_{aIJ}$ is the connection compatible with $\tilde{\Pi}^{aI}$~\eqref{CDP}. 
	
	Therefore, using the parametrization of $e^I$ and the spatial part of the connection~\eqref{w3d}---together with the correspondent redefinition of the Lagrange multiplier accompanying the Gauss constraint---the action for the theory in three dimensions is 
	\begin{eqnarray}
		\label{S_Q_3d}
		S & = & \int_{\mathbb{R} \times \Sigma} dtd^{2}x \bigg[  2 \kappa \tilde{\Pi}^{aI} \dot{Q}_{aI} + \dfrac{1}{2} h^{1/2} n_{I} \Big( \bar{\psi} \gamma^{I} E \dot{\psi}  \notag \\
		&&-  \dot{\bar{\psi}} \gamma^{I} E^{\dagger} \psi \Big) - \lambda_{IJ} \tilde{\mathcal{G}}^{IJ} - 2 N^{a} \tilde{\mathcal{D}}_{a} - \uac{N} \tilde{\tilde{\mathcal{H}}} \bigg].
	\end{eqnarray}
	Here, the coupling matrix is $E=(1 + \mathrm{i} \theta ) \mathds{1}$ and the first-class constraints $\tilde{\mathcal{G}}^{IJ}$, $\tilde{\mathcal{D}}_{a}$, and  $\tilde{\tilde{\mathcal{H}}}$ are given by
	\begin{widetext}
		\begin{subequations} 
			\label{const_Q_3d}
			\begin{eqnarray}
				\tilde{\mathcal{G}}^{IJ} & = & 2 \kappa \tilde{\Pi}^{a[I} Q_{a}{}^{J]}
				+ \dfrac{\theta}{2} h^{1/2} n^{[I} V^{J]} + \dfrac{1}{4} h^{1/2} \epsilon^{IJK} n_{K} \bar{\psi} \psi , \\
				\tilde{\mathcal{D}}_{a} & = & \kappa \Big( 2 \tilde{\Pi}^{bI} \partial_{[a} Q_{b]I} - Q_{aI} \partial_{b} \tilde{\Pi}^{bI} \Big) + \frac{1}{4} h^{1/2} n_{I} \left(\bar{\psi} \gamma^{I} \partial_{a} \psi - \partial_{a} \bar{\psi} \gamma^{I} \psi + \theta \partial_{a} V^{I} \right), \\
				\tilde{\tilde{\mathcal{H}}} & = & \kappa \tilde{\Pi}^{aI} \tilde{\Pi}^{bJ} R_{abIJ} + 2 \kappa \tilde{\Pi}^{a[I} \tilde{\Pi}^{|b|J]} Q_{aI} Q_{bJ} + \dfrac{1}{2} h^{1/2} \tilde{\Pi}^{aI} \left( \bar{\psi} \gamma_{I} \nabla_{a} \psi - \overline{\nabla_{a} \psi} \gamma_{I} \psi \right) \notag \\
				&& + \dfrac{1}{2} h^{1/2} n_{I} Q_{aJ}  \left( \theta \tilde{\Pi}^{aJ} V^{I} - \epsilon^{IJK} \tilde{\Pi}^{a}{}_{K} \bar{\psi} \psi \right) + \dfrac{h}{8 \kappa} \theta^{2} q_{IJ} V^{I} V^{J}  - h \left( 2 \kappa \Lambda + m \bar{\psi} \psi  \right),	
			\end{eqnarray}
		\end{subequations}
	\end{widetext}
	where $V^{I}$ is defined in~\eqref{Vc_def}, and~\eqref{id_anticomm} for $n=3$ is also used, which amounts to 
	\begin{equation}
		\label{id_3d}
		\{\gamma^{I}, \sigma^{JK}\} =  \epsilon^{IJK} \mathds{1},
	\end{equation}
	because of $\Gamma = - \gamma^{0} \gamma^{1} \gamma^{2} = \mathds{1}$ (and therefore, $[\gamma^{I}, \gamma^{J}] = 2 \epsilon^{IJ}{}_{K} \gamma^{K}$).
	
	Note that the Hamiltonian formulation~\eqref{S_Q_3d} is the same one given in~\eqref{S_Q} by substituting $n=3$ and \eqref{id_3d} in~\eqref{const_Q_G_o}--\eqref{const_Q_H_o}. Therefore, all the Hamiltonian formulations obtained after~\eqref{S_Q} also hold for a spacetime of dimension three.
	
	For the sake of completeness, we give two additional Hamiltonian formulations that are particular of a spacetime of dimension three.

	\subsection{First formulation} 
	
	The first of them is obtained via a symplectomorphism that only replaces the variable $Q_{aI}$ with the variable 
	\begin{equation}
		\label{Q2A}
		A_{aI} = \dfrac{1}{2} \epsilon_{IJK} \Gamma_{a}{}^{JK} + Q_{aI},
	\end{equation}
	and leaves $\tilde{\Pi}^{aI}$, $\psi$, and $\bar{\psi}$ unchanged. Since $\Gamma_{aIJ}$ is a connection and $Q_{aI}$ is a vector, the variable $A_{aI}$ is also a connection. In terms of the new phase-space variables---and neglecting a boundary term---the action~\eqref{S_Q_3d} becomes 
	\begin{eqnarray}
		\label{S_Ash_3d}
		S & = & \int_{\mathbb{R} \times \Sigma} dtd^{2}x \bigg[  2 \kappa \tilde{\Pi}^{aI} \dot{A}_{aI} + \dfrac{1}{2} h^{1/2} n_{I} \Big( \bar{\psi} \gamma^{I} E \dot{\psi}  \notag \\
		&&-  \dot{\bar{\psi}} \gamma^{I} E^{\dagger} \psi \Big) - \lambda_{I} \tilde{\mathcal{G}}^{I} - 2 N^{a} \tilde{\mathcal{D}}_{a} - \uac{N} \tilde{\tilde{\mathcal{H}}} \bigg],
	\end{eqnarray}
	where the first-class constraints are given by
	\begin{widetext}
		\begin{subequations}
			\begin{eqnarray}
				\label{const_Ash_G}
				\tilde{\mathcal{G}}^{I} & := &  - \dfrac{1}{2} \epsilon^{IJK} \tilde{\mathcal{G}}_{JK} =  \kappa \left( \partial_{a} \tilde{\Pi}^{aI} + \epsilon^{I}{}_{JK} A_{a}{}^{J} \tilde{\Pi}^{aK} \right) - \dfrac{\theta}{4} h^{1/2} \epsilon^{IJK} n_{J} V_{K} + \dfrac{1}{4} h^{1/2}  n^{I} \bar{\psi} \psi , \\
				\tilde{\mathcal{D}}_{a} & = & \kappa \Big( 2 \tilde{\Pi}^{bI} \partial_{[a} A_{b]I} - A_{aI} \partial_{b} \tilde{\Pi}^{bI} \Big) + \frac{1}{4} h^{1/2} n_{I} \left(\bar{\psi} \gamma^{I} \partial_{a} \psi - \partial_{a} \bar{\psi} \gamma^{I} \psi + \theta \partial_{a} V^{I} \right), \\
				\tilde{\tilde{\mathcal{H}}} & = & - \kappa \epsilon_{IJK} \tilde{\Pi}^{aI} \tilde{\Pi}^{bJ} F_{ab}{}^{K} + \dfrac{1}{2} h^{1/2} \tilde{\Pi}^{aI} \left( \bar{\psi} \gamma_{I} \nabla_{a} \psi - \overline{\nabla_{a} \psi} \gamma_{I} \psi \right) + \dfrac{1}{2} h^{1/2} n_{I} \Big( A_{aJ} - \dfrac{1}{2} \epsilon_{JKL} \Gamma_{a}{}^{KL} \Big)  \Big( \theta \tilde{\Pi}^{aJ} V^{I} \notag \\
				&& - \epsilon^{IJM} \tilde{\Pi}^{a}{}_{M} \bar{\psi} \psi \Big) - \dfrac{\theta}{2} h^{1/2} \epsilon_{IJK} \tilde{\Pi}^{aI} n^{J} \nabla_{a} V^{K} + \dfrac{h}{8 \kappa} \theta^{2} q_{IJ} V^{I} V^{J}  - h \left( 2 \kappa \Lambda + m \bar{\psi} \psi  \right),
			\end{eqnarray}
		\end{subequations}
		\newpage
	\end{widetext}
	with $F_{abI} = \partial_{a} A_{bI} - \partial_{b} A_{aI} + \epsilon_{IJK} A_{a}{}^{J} A_{b}{}^{K}$ being the curvature of $A_{aI}$, and we have used the identity \begin{eqnarray}
		2 \tilde{\Pi}^{aI}\nabla_{a} \tilde{\mathcal{G}}_{I} &=& - \kappa \tilde{\Pi}^{aI} \tilde{\Pi}^{bJ} \big( R_{abIJ} + \epsilon_{IJK} F_{ab}{}^{K} \big)
		\notag \\
		&& - 2 \kappa \tilde{\Pi}^{a[I} \tilde{\Pi}^{|b|J]} \Big( A_{aI}  - \dfrac{1}{2} \epsilon_{IKL}\Gamma_{a}{}^{KL} \Big)    \notag\\
		&&  \times \Big( A_{bJ}  - \dfrac{1}{2} \epsilon_{JMN}\Gamma_{b}{}^{MN} \Big) \notag \\
		&& - \dfrac{\theta}{2} h^{1/2} \epsilon_{IJK} \tilde{\Pi}^{aI} n^{J} \nabla_{a} V^{K}, 
	\end{eqnarray}
	and redefined the Lagrange multiplier by  $\lambda_{I} :=   \epsilon_{IJK} \lambda^{JK} + 2 \tilde{\Pi}^{a}{}_{I} \nabla_{a} \uac{N}$.
	
	\subsection{Second formulation}
	
	Similarly to the analysis made in Sec.~\ref{Sec_2more}, we define the half-densitized fermions
	\begin{subequations}
		\begin{eqnarray}
			\phi & := & h^{1/4} \psi, \\
			\bar{\phi} & := & h^{1/4} \bar{\psi},
		\end{eqnarray}
	\end{subequations}
	and the variable 
	\begin{eqnarray}
		\label{A2L}
		\mathcal{A}_{aI} &:=& A_{aI} + \dfrac{\mathrm{i} \theta}{2 \kappa}  \uacc{h}_{ab} \tilde{\Pi}^{b}{}_{[I} n_{J]} \bar{\phi} \gamma^{J} \phi \notag \\
		& = & A_{aI} + \dfrac{\theta}{2 \kappa}  \uacc{h}_{ab} \tilde{\Pi}^{b}{}_{[I} n_{J]} \tilde{V}^{J},
	\end{eqnarray}
	where the densitized vector current $\tilde{V}^{I}$ is defined by~\eqref{Vcd_def}. Furthermore, since $\tilde{\Pi}^{aI}$, $n_{I}$, and $\tilde{V}^{I}$ are Lorentz vectors, the second term on the right-hand-side of \eqref{A2L} transforms as a Lorentz tensor. Therefore, $\mathcal{A}_{aI}$ is a Lorentz connection. Then, in terms of these phase-space variables, the action~\eqref{S_Ash_3d} acquires the form
	\begin{eqnarray}
		S & = & \int_{\mathbb{R} \times \Sigma} dtd^{2}x \bigg[  2 \kappa \tilde{\Pi}^{aI} \dot{\mathcal{A}}_{aI} + \dfrac{1}{2} n_{I} \Big( \bar{\phi} \gamma^{I} \dot{\phi}  \notag \\
		&&-  \dot{\bar{\phi}} \gamma^{I}  \phi \Big) - \lambda_{I} \tilde{\mathcal{G}}^{I} - 2 N^{a} \tilde{\mathcal{D}}_{a} - \uac{N} \tilde{\tilde{\mathcal{H}}} \bigg],
	\end{eqnarray}
	where the first-class constraints are given by
	\begin{widetext}
		\begin{subequations}
			\begin{eqnarray}
				\tilde{\mathcal{G}}^{I} & = &   \kappa \left( \partial_{a} \tilde{\Pi}^{aI} + \epsilon^{I}{}_{JK} \mathcal{A}_{a}{}^{J} \tilde{\Pi}^{aK} \right)  + \dfrac{1}{4} n^{I} \bar{\phi} \phi , \\
				\tilde{\mathcal{D}}_{a} & = & \kappa \Big( 2 \tilde{\Pi}^{bI} \partial_{[a} \mathcal{A}_{b]I} - \mathcal{A}_{aI} \partial_{b} \tilde{\Pi}^{bI} \Big) + \frac{1}{4}  n_{I} \left(\bar{\phi} \gamma^{I} \partial_{a} \phi - \partial_{a} \bar{\phi} \gamma^{I} \phi  \right), \\
				\tilde{\tilde{\mathcal{H}}} & = & - \kappa \epsilon_{IJK} \tilde{\Pi}^{aI} \tilde{\Pi}^{bJ} \mathcal{F}_{ab}{}^{K} + \dfrac{1}{2} \tilde{\Pi}^{aI} \left( \bar{\phi} \gamma_{I} \nabla_{a} \phi - \overline{\nabla_{a} \phi} \gamma_{I} \phi \right)  \notag \\
				&&  - \dfrac{1}{2} \epsilon_{IJK} n^{I} \Big( \mathcal{A}_{a}{}^{J} - \dfrac{1}{2} \epsilon^{JLM} \Gamma_{aLM} \Big) \tilde{\Pi}^{aK}  \bar{\phi} \phi  + \dfrac{1}{8 \kappa} \theta^{2}  \tilde{V}^{I} \tilde{V}_{I}  \notag \\
				&& -  2 \kappa h \Lambda -  h^{1/2} m \bar{\phi} \phi ,
			\end{eqnarray}
		\end{subequations}
	\end{widetext}
	with $\mathcal{F}_{abI} = \partial_{a} \mathcal{A}_{bI} - \partial_{b} \mathcal{A}_{aI} + \epsilon_{IJK} \mathcal{A}_{a}{}^{J} \mathcal{A}_{b}{}^{K}$ being the curvature of $\mathcal{A}_{aI}$. This Hamiltonian formulation also has the
	peculiarity that the coupling parameter $\theta$ only appears in the quartic-fermion interaction. Therefore, if the coupling of fermions is minimal (and thus $\theta = 0$), there are no quartic fermion interactions.

	\section{Hamiltonian formulations when $n=4$}\label{Appendixn=4}
	The Hamiltonian formulations presented in Secs.~\ref{Sec_HA} and \ref{Sec_2more} are valid for $n\geq 3$. Here, we restrict the analysis to $n=4$ and use~\eqref{id_4d} to rewrite the formulations given by the actions~\eqref{S_Q}, \eqref{S_hf}, and \eqref{S_hf_2}.

	\subsection{First formulation} Thus, if $n=4$, the first-class constraints of the action~\eqref{S_Q} acquire the form
		\begin{widetext}
		\begin{subequations} 
			\label{const_Q_4d}
			\begin{eqnarray}
				\label{const_Q_G_4d}
				\tilde{\mathcal{G}}^{IJ} & = & 2 \kappa \tilde{\Pi}^{a[I} Q_{a}{}^{J]}
				+ \dfrac{1}{2} h^{1/4} n^{[I} \left( \theta V^{J]}  + \xi A^{J]} \right) + \dfrac{1}{4} h^{1/4} \epsilon^{IJKL}  n_{K} A_{L} , \\
				\label{const_Q_D_4d}
				\tilde{\mathcal{D}}_{a} & = & \kappa \Big( 2 \tilde{\Pi}^{bI} \partial_{[a} Q_{b]I} - Q_{aI} \partial_{b} \tilde{\Pi}^{bI} \Big) + \frac{1}{4} h^{1/4} n_{I}  \left[\bar{\psi} \gamma^{I} \partial_{a} \psi - \partial_{a} \bar{\psi} \gamma^{I} \psi + \partial_{a}\left( \theta V^{I} + \xi A^{I} \right) \right], \\
				\tilde{\tilde{\mathcal{H}}} & = & \kappa \tilde{\Pi}^{aI} \tilde{\Pi}^{bJ} R_{abIJ} + 2 \kappa \tilde{\Pi}^{a[I} \tilde{\Pi}^{|b|J]} Q_{aI} Q_{bJ} + \dfrac{1}{2} h^{1/4} \tilde{\Pi}^{aI} \left( \bar{\psi} \gamma_{I} \nabla_{a} \psi - \overline{\nabla_{a} \psi} \gamma_{I} \psi \right) \notag \\
				&& + \dfrac{1}{2} h^{1/4} n_{I} Q_{aJ}  \left[ \tilde{\Pi}^{aJ}  \left( \theta V^{I} + \xi A^{I} \right) -  \epsilon^{IJKL} \tilde{\Pi}^{a}{}_{K} A_{L} \right] + \dfrac{3h^{1/2}}{32 \kappa} \Big[ n_{I} n_{J} A^{I} A^{J} \notag \\
				&&  + q_{IJ} \left( \theta^{2} V^{I} V^{J} + \xi^{2} A^{I} A^{J} + 2 \theta \xi V^{I} A^{J} \right)  \Big]   - h^{1/2} \left( 2 \kappa \Lambda + m \bar{\psi} \psi  \right),
			\end{eqnarray}
		\end{subequations}
	where we have used~\eqref{id_4d} and the fermion currents $V^{I}$ and $A^{I}$ are defined in~\eqref{Vc_def} and~\eqref{Ac_def}, respectively. 
	
	\subsection{Second formulation}
	
	Similarly, if $n=4$ the first-class constraints of the Hamiltonian formulation~\eqref{S_hf} become
		\begin{subequations}
			\begin{eqnarray}
				\tilde{\mathcal{G}}^{IJ} & = & 2 \kappa \tilde{\Pi}^{a[I} \Psi_{a}{}^{J]}
				+ \dfrac{1}{4} \epsilon^{IJKL} n_{K} \tilde{A}_{L} , \\
				\tilde{\mathcal{D}}_{a} & = & \kappa \Big( 2 \tilde{\Pi}^{bI} \partial_{[a} \Psi_{b]I} - \Psi_{aI} \partial_{b} \tilde{\Pi}^{bI} \Big) + \frac{1}{4} n_{I}  \left( \bar{\phi} \gamma^{I}  \partial_{a} \phi - \partial_{a} \bar{\phi} \gamma^{I} \phi \right), \\
				\tilde{\tilde{\mathcal{H}}} & = & \kappa \tilde{\Pi}^{aI} \tilde{\Pi}^{bJ} R_{abIJ} + 2 \kappa \tilde{\Pi}^{a[I} \tilde{\Pi}^{|b|J]} \Psi_{aI} \Psi_{bJ} + \dfrac{1}{2} \tilde{\Pi}^{aI} \left( \bar{\phi} \gamma_{I} \nabla_{a} \phi - \overline{\nabla_{a} \phi} \gamma_{I} \phi \right) \notag \\
				&& + \dfrac{1}{2} \epsilon_{IJKL} n^{I} \tilde{\Pi}^{aJ} \Psi_{a}{}^{K} \tilde{A}^{L} + \dfrac{3}{32 \kappa} \Big[   \theta^{2} \tilde{V}^{I} \tilde{V}_{I} + 2 \theta \xi \tilde{V}^{I} \tilde{A}_{I} + \xi^{2} \tilde{A}^{I} \tilde{A}_{I} +  n_{I} n_{J} \tilde{A}^{I} \tilde{A}^{J}\Big]  \notag \\
				&&  - 2 h^{1/2} \kappa \Lambda - h^{1/4} m \bar{\phi} \phi,		
			\end{eqnarray}
		\end{subequations}
	\end{widetext}
	where $\tilde{V}^{I}$ and $\tilde{A}^{I}$ are the densitized fermion currents defined in~\eqref{Vcd_def} and \eqref{Acd_def}.  
	
	Remark: Note that is possible to perform a symplectomorphism from this formulation and obtain the Hamiltonian formulation of fermions coupled to the Holst action reported in the Eq. (64) of Ref.~\cite{Romero2106}. The symplectomorphism implies to change the gravitational variables $\Psi_{aI}$ with 
	\begin{equation}
		\label{P2H}
		\varphi_{aI} := \Psi_{aI} + W_{a}{}^{b}{}_{IJK} \left( \delta^{[J}_{M} \delta^{K]}_{N} + \dfrac{1}{2\gamma} \epsilon^{JK}{}_{MN} \right) \Gamma_{b}{}^{MN},
	\end{equation}
	and leave $\tilde{\Pi}^{aI}$, $\phi$, and $\bar{\phi}$ unchanged. By doing so, and also choosing the parameters involved in this Hamiltonian formulation as
	\begin{subequations}
		\begin{eqnarray}	
			\theta &=& \dfrac{\gamma}{\sqrt{\gamma^{2} +1}} \vartheta, \\
			\xi &=& \dfrac{\gamma}{\sqrt{\gamma^{2} +1}} \left(\zeta + \dfrac{1}{\gamma} \right),
		\end{eqnarray}
	\end{subequations}
	where $\vartheta$, $\zeta \in \mathbb{R}$ and $\gamma$ is the Barbero-Immirzi parameter, we get precisely the Hamiltonian formulation given in Eq. (64) of Ref.~\cite{Romero2106}.	
	
	\subsection{Third formulation}
	
	If $n=4$, then the first-class constraints of the Hamiltonian formulation~\eqref{S_hf_2} acquire the form 
	\begin{widetext}
		\begin{subequations}
			\begin{eqnarray}
				\label{const_hf_G_4d}
				\tilde{\mathcal{G}}^{IJ} & = & 2 \kappa \tilde{\Pi}^{a[I} \Psi_{a}{}^{J]}
				+ \dfrac{1}{4} \epsilon^{IJKL} n_{K} \tilde{A}_{L}, \\
				\label{const_hf_D_4d}
				\tilde{\mathcal{D}}_{a} & = & \kappa \Big( 2 \tilde{\Pi}^{bI} \partial_{[a} \Psi_{b]I} - \Psi_{aI} \partial_{b} \tilde{\Pi}^{bI} \Big) + \frac{1}{4} n_{I}  \left( \bar{\phi} \gamma^{I}  \partial_{a} \phi - \partial_{a} \bar{\phi} \gamma^{I} \phi \right), \\
				\label{const_hf_C_4d}
				\tilde{\tilde{\mathcal{C}}} & = & \kappa \tilde{\Pi}^{aI} \tilde{\Pi}^{bJ} R_{abIJ} + 2 \kappa \tilde{\Pi}^{a[I} \tilde{\Pi}^{|b|J]} \Psi_{aI} \Psi_{bJ} + \dfrac{1}{2} \tilde{\Pi}^{aI} \left( \bar{\phi} \gamma_{I} \nabla_{a} \phi - \overline{\nabla_{a} \phi} \gamma_{I} \phi \right) \notag \\
				&&  + \dfrac{1}{32 \kappa} \left\lbrace   3 \left[ \theta^{2} \tilde{V}_{I} \tilde{V}^{I} + 2 \theta \xi \tilde{V}_{I} \tilde{A}^{I} +  \left( \xi^{2} - 1 \right) \tilde{A}_{I} \tilde{A}^{I} \right] - q_{IJ} \tilde{A}^{I} \tilde{A}^{J} \right\rbrace  \notag \\
				&& - 2 h^{1/2} \kappa \Lambda - h^{1/4} m \bar{\phi} \phi .
			\end{eqnarray}
		\end{subequations}
	
	Using the identity
	\begin{eqnarray}
		\label{id_qAA}
		q_{IJ} \tilde{A}^{I} \tilde{A}^{J} &=&  - 4 \kappa \epsilon_{IJKL} n^{I} \tilde{\Pi}^{aJ} \Psi_{a}{}^{K} \tilde{A}^{L}  + 2 \epsilon_{IJKL} \tilde{\mathcal{G}}^{IJ} n^{K} \tilde{A}^{L},
	\end{eqnarray}
	the Hamiltonian constraint \eqref{const_hf_C_4d} acquires the form
		\begin{eqnarray}
			\tilde{\tilde{\mathcal{C}}} & = & \kappa \tilde{\Pi}^{aI} \tilde{\Pi}^{bJ} R_{abIJ} + 2 \kappa \tilde{\Pi}^{a[I} \tilde{\Pi}^{|b|J]} \Psi_{aI} \Psi_{bJ} + \dfrac{1}{2} \tilde{\Pi}^{aI} \left( \bar{\phi} \gamma_{I} \nabla_{a} \phi - \overline{\nabla_{a} \phi} \gamma_{I} \phi \right) \notag \\
			&&  + \dfrac{1}{8} \epsilon_{IJKL} n^{I} \tilde{\Pi}^{aJ} \Psi_{a}{}^{K} \tilde{A}^{L} + \dfrac{3}{32 \kappa}  \left[ \theta^{2} \tilde{V}_{I} \tilde{V}^{I} + 2 \theta \xi \tilde{V}_{I} \tilde{A}^{I} +  \left( \xi^{2} - 1 \right) \tilde{A}_{I} \tilde{A}^{I} \right]   \notag \\
			&& - 2 h^{1/2} \kappa \Lambda - h^{1/4} m \bar{\phi} \phi - \dfrac{1}{16 \kappa}\epsilon_{IJKL}  \tilde{\mathcal{G}}^{IJ} n^{K} \tilde{A}^{L},
		\end{eqnarray}
		and, finally, factoring out the Gauss constraint, the Hamiltonian constraint becomes
		\begin{eqnarray}
		    \label{const_hf_H_4d_}
			\tilde{\tilde{H}} & := & \kappa \tilde{\Pi}^{aI} \tilde{\Pi}^{bJ} R_{abIJ} + 2 \kappa \tilde{\Pi}^{a[I} \tilde{\Pi}^{|b|J]} \Psi_{aI} \Psi_{bJ} + \dfrac{1}{2} \tilde{\Pi}^{aI} \left( \bar{\phi} \gamma_{I} \nabla_{a} \phi - \overline{\nabla_{a} \phi} \gamma_{I} \phi \right) \notag \\
			&&  + \dfrac{1}{8} \epsilon_{IJKL} n^{I} \tilde{\Pi}^{aJ} \Psi_{a}{}^{K} \tilde{A}^{L} + \dfrac{3}{32 \kappa}  \left[ \theta^{2} \tilde{V}_{I} \tilde{V}^{I} + 2 \theta \xi \tilde{V}_{I} \tilde{A}^{I} +  \left( \xi^{2} - 1 \right) \tilde{A}_{I} \tilde{A}^{I} \right]   \notag \\
			&& - 2 h^{1/2} \kappa \Lambda - h^{1/4} m \bar{\phi} \phi.
		\end{eqnarray}

	Therefore, the resulting Hamiltonian formulation is given by the action
	\begin{eqnarray}
		\label{S_hf_4d}
		S & = & \int_{\mathbb{R} \times \Sigma} dtd^{3}x \bigg[  2 \kappa \tilde{\Pi}^{aI} \dot{\Psi}_{aI} + \dfrac{1}{2} n_{I} \Big( \bar{\phi} \gamma^{I} \dot{\phi}  - \dot{\bar{\phi}} \gamma^{I}  \phi \Big)  \notag \\
		&& - \nu_{IJ} \tilde{\mathcal{G}}^{IJ} - 2 N^{a} \tilde{\mathcal{D}}_{a} - \uac{N} \tilde{\tilde{H}} \bigg],
	\end{eqnarray}
	with the constraints~\eqref{const_hf_G_4d}, \eqref{const_hf_D_4d}, and \eqref{const_hf_H_4d_}. This formulation is relevant because it allows us to see how the Hamiltonian formulation of the particular Lagrangian action~\eqref{S} that is equivalent to the Einstein-Dirac theory looks like. We recall the reader that the Lagrangian action~\eqref{S} is equivalent to the Einstein-Dirac theory if the parameters in the matrix coupling are chosen such that $E =\mathds{1} - \tau \mathrm{i} \Gamma$  (which amounts to set $\theta=0$ and $\xi=\tau$, see Sec.~\ref{ss_second_order_4d}). Thus, for this particular coupling, the first-class constraints are given by~\eqref{const_hf_G_4d}, \eqref{const_hf_D_4d}, and \eqref{const_hf_H_4d_} acquires the form
		\begin{subequations}
			\begin{eqnarray}
				\tilde{\tilde{H}} & = & \kappa \tilde{\Pi}^{aI} \tilde{\Pi}^{bJ} R_{abIJ} + 2 \kappa \tilde{\Pi}^{a[I} \tilde{\Pi}^{|b|J]} \Psi_{aI} \Psi_{bJ} + \dfrac{1}{2} \tilde{\Pi}^{aI} \left( \bar{\phi} \gamma_{I} \nabla_{a} \phi - \overline{\nabla_{a} \phi} \gamma_{I} \phi \right) \notag \\
				&&  + \dfrac{1}{8} \epsilon_{IJKL} n^{I} \tilde{\Pi}^{aJ} \Psi_{a}{}^{K} \tilde{A}^{L}  - 2 h^{1/2} \kappa \Lambda - h^{1/4} m \bar{\phi} \phi.
			\end{eqnarray}
		\end{subequations}
	Note that the resulting Hamiltonian description does not involve any quartic-fermion interactions.
	
	This shows, by the way, that it is also possible to get a Hamiltonian formulation without quartic-fermion interactions for the Holst action from each one of the half-densitized fermion formulations presented in Ref.~\cite{Romero2106} using an analogous identity to~\eqref{id_qAA} and choosing the appropriate coupling parameters (see footnote $2$ of Sec.~\ref{Sec_Lagran}).
	\end{widetext}

	\bibliographystyle{apsrev4-1}
	
	\bibliography{References}
	
\end{document}